\documentclass{mn2e}
\pdfoutput=1

\usepackage{graphicx,txfonts,amssymb,amsfonts,times}
\usepackage{natbib}
\usepackage{subfig}
\usepackage{epsfig}
\usepackage{color,float}
\usepackage[usenames,dvipsnames,svgnames,table]{xcolor}
\usepackage{enumerate}
\usepackage{epstopdf}
\usepackage{fixltx2e}
\usepackage{booktabs}
\DeclareGraphicsExtensions{.pdf,.png,.jpeg,.ps,.eps}
\usepackage{epsfig}
\usepackage{threeparttable}
\usepackage{framed}
\usepackage[T1]{fontenc}
\usepackage{aecompl}
\usepackage{soul}

\begin{document}
\newcommand{\note}[1]{\textbf{\textcolor{red}{{\fontfamily{garamond}\selectfont{#1}}}}}
\newcommand{\lclash}{Cluster Lensing And Supernova survey with Hubble}
\newcommand{\kms}{~km~s$^{-1}$}
\newcommand{\logh}{$+5\log h_{70}$}
\newcommand{\ho}{$\sim$}
\newcommand{\hi}{$h^{-1}_{70}$~}
\newcommand{\msun}{M$_\odot$~}
\newcommand{\degtwo}{deg$^2$}
\newcommand{\sbu}{mag arcsec$^{-2}$}
\newcommand{\firstpaper}{(Paper I)}
\newcommand{\spitzer}{\textit{Spitzer}}
\newcommand{\galex}{{\it GALEX}}
\newcommand{\herschel}{\textit{Herschel}}
\newcommand{\hst}{\textit{HST}}
\newcommand{\chandra}{\textit{Chandra}}
\newcommand{\irac}{IRAC}
\newcommand{\irs}{IRS}
\newcommand{\mips}{MIPS}
\newcommand{\wfc}{WFC3}
\newcommand{\acs}{ACS}
\newcommand{\galfit}{\texttt{GALFIT}}
\newcommand{\multidrizzle}{\texttt{MultiDrizzle}}
\newcommand{\ergflux}{~erg~s$^{-1}$~cm$^{-2}$}
\newcommand{\mfive}{M$_{500}$}
\newcommand{\mtwo}{M$_{200}$}
\newcommand{\rfive}{r$_{500}$}
\newcommand{\rtwo}{r$_{200}$}
\newcommand{\se}{Source Extractor}
\newcommand{\blue}{F110W}
\newcommand{\red}{F160W}
\newcommand{\green}{F125W}
\newcommand{\til}{$\sim$}
\newcommand{\MACStwelve}{MACS1206.2-0847}
\newcommand{\MACSohthree}{MACS0329.7-0211}
\newcommand{\MACSeleven}{MACS1149.6+2223}
\newcommand{\MACStwentyone}{MACS2129.4-0741}
\newcommand{\Abell}{Abell 2261}
\newcommand{\zform}{z$_{f}$}
\newcommand{\Lstar}{L*}
\newcommand{\Zsolar}{Z$_\odot$}
\newcommand{\Msun}{M$_\odot$}
\newcommand{\Mstar}{M$_*$}
\newcommand{\metgrad}{d(Z/Z$_\odot$)(d $\log$(r))$^{-1}$}
\newcommand{\colorgrad}{d(F110W-F160W)(d $\log$(r))$^{-1}$}

\def\aj{\rmfamily{AJ}}
\def\apj{\rmfamily{ApJ}}
\def\apjl{\rmfamily{ApJ}}
\def\apjs{\rmfamily{ApJS}}
\let\apjlett=\apjl
\let\apjsupp=\apjs
\def\aap{\rmfamily{A\&A}}
\let\astap=\aap
\def\araa{\rmfamily{ARA\&A}}
\def\aapr{\rmfamily{A\&A~Rev.}}
\def\aaps{\rmfamily{A\&AS}}
\def\mnras{\rmfamily{MNRAS}}
\def\pasp{\rmfamily{PASP}}
\def\apss{\rmfamily{Ap\&SS}}
\def\pasj{\rmfamily{PASJ}}
\def\pra{\rmfamily{Phys.~Rev.~A}}
\def\prb{\rmfamily{Phys.~Rev.~B}}
\def\prc{\rmfamily{Phys.~Rev.~C}}
\def\prd{\rmfamily{Phys.~Rev.~D}}
\def\pre{\rmfamily{Phys.~Rev.~E}}
\def\rmp{\rmfamily{Rev.~Mod.~Phys.}}
\def\prl{\rmfamily{Phys.~Rev.~Lett.}}
\def\nat{\rmfamily{Nature}}
\def\sci{\rmfamily{Science}}
\def\skytel{\rmfamily{S\&T}}
\def\ao{\rmfamily{Appl.~Opt.}}
\let\applopt=\ao
\def\azh{\rmfamily{AZh}}
\def\baas{\rmfamily{BAAS}}
\def\jrasc{\rmfamily{JRASC}}
\def\memras{\rmfamily{MmRAS}}
\def\qjras{\rmfamily{QJRAS}}
\def\solphys{\rmfamily{Sol.~Phys.}}
\def\sovast{\rmfamily{Soviet~Ast.}}
\def\ssr{\rmfamily{Space~Sci.~Rev.}}
\def\zap{\rmfamily{ZAp}}
\def\iauc{\rmfamily{IAU~Circ.}}
\let\iaucirc=\iauc
\def\aplett{\rmfamily{Astrophys.~Lett.}}
\def\apspr{\rmfamily{Astrophys.~Space~Phys.~Res.}}
\def\bain{\rmfamily{Bull.~Astron.~Inst.~Netherlands}}
\def\fcp{\rmfamily{Fund.~Cosmic~Phys.}}
\def\gca{\rmfamily{Geochim.~Cosmochim.~Acta}}
\def\grl{\rmfamily{Geophys.~Res.~Lett.}}
\def\jcp{\rmfamily{J.~Chem.~Phys.}}
\def\jgr{\rmfamily{J.~Geophys.~Res.}}
\def\jqsrt{\rmfamily{J.~Quant.~Spec.~Radiat.~Transf.}}
\def\memsai{\rmfamily{Mem.~Soc.~Astron.~Italiana}}%
\def\nphysa{\rmfamily{Nucl.~Phys.~A}}
\def\physrep{\rmfamily{Phys.~Rep.}}
\def\physscr{\rmfamily{Phys.~Scr}}
\def\planss{\rmfamily{Planet.~Space~Sci.}}
\def\procspie{\rmfamily{Proc.~SPIE}}
\def\aipconf{\rmfamily{AIP~Conf.~Proc.}}
\def\aspconf{\rmfamily{ASP~Conf.~Ser.}}
\def\asslconf{\rmfamily{Ap\&SSL~Conf.~Ser.}}
\def\maps{\rmfamily{MAPS}}


\title[On the Origin of the Intracluster Light in Massive Galaxy Clusters]
{On the Origin of the Intracluster Light in Massive Galaxy Clusters}

\author[T. De Maio et al. 2014]
{Tahlia De Maio$^1$, Anthony Gonzalez$^1$, Ann Zabludoff $^2$, Dennis Zaritsky $^2$, Maru\u{s}a Brada\u{c},$^3$
  \\
   $^1$ Department of Astronomy, University of Florida, Gainesville, FL 32611\\
  $^2$ Department of Astronomy, University of Arizona, Steward Observatory, Tucson, AZ  85721 \\
  $^3$ Department of Physics, University of California- Davis, Davis, CA 95616 \\
}

\maketitle

\begin{abstract}
We present a pilot study on the origin and assembly history of the ICL for four galaxy clusters at $0.44\le z\le0.57$ observed with the Hubble Space Telescope from the Cluster Lensing and Supernova Survey with Hubble (CLASH) sample.  
Using this sample of CLASH clusters we set an empirical limit on the amount of scatter in ICL surface brightness profiles of such clusters at z=0.5, a mean of 0.24 \sbu\ for 10$<$r$<$110 kpc, and constrain the progenitor population and formation mechanism of the ICL by measuring the ICL surface brightness profile, the ICL colour and colour gradient, and the total ICL luminosity within the same radial range.  
This scatter is physical --  it exceeds the observational errors, straightforward expectations from the range of cluster masses in our sample, and predictions based on published evolutionary models for the variance attributable to the redshift span of our sample.
We associate the additional scatter with differences in ICL assembly process, formation epoch, and/or ICL content.
Using stellar population synthesis models we transform the observed colours to metallicity.
For three of the four clusters we find clear negative gradients that, on average, decrease from super solar in the central regions of the Brightest Cluster Galaxy (BCG) to sub-solar in the ICL, under the assumption that the age of the intracluster stars is $>$11 Gyrs.
Such negative colour (and equivalently, metallicity) gradients can arise from tidal stripping of L* galaxies and/or the disruption of dwarf galaxies, but not major mergers with the BCG.
We also find that the ICL at 110 kpc has a colour comparable to m*+2 red sequence galaxies, suggesting that out to this radius the ICL is dominated by stars liberated from galaxies with L$>$0.2 \Lstar.
Finally, we find ICL luminosities of 4-8 \Lstar\ between 10$<$r$<$110 kpc for these clusters. 
Neither dwarf disruption or major mergers with the BCG alone can explain this level of luminosity and remain consistent with either the observed evolution in the faint end slope of the luminosity function or predictions for the number of BCG major mergers since z=1.
Taken together, the results of this pilot study are suggestive of a formation history for these clusters in which the ICL is built-up by the stripping of >0.2\Lstar\ galaxies, and disfavour significant contribution to the ICL by dwarf disruption or major mergers with the BCG.

\end{abstract}


\section{Introduction}     
\label{sec:intro}

Galaxy clusters lie at the crossroads of astrophysics and cosmology, probing the influence of dark matter and dark energy in the evolution of large-scale structures in the universe. 
There exists a large body of research on cluster galaxies, the intracluster medium (ICM), and dark matter (\cite{Kravtsov2012a}, references therein). 
The properties of the intracluster starlight (ICL), the light from stars bound to the cluster potential but not to individual galaxies, remain less well-determined than those of the other cluster components. 
In recent years, several deep ICL surveys of nearby clusters have been completed \citep{Gonzalez2000, Feldmeier2002, Feldmeier2004a, Gonzalez2005,  KrickI, KrickII}. 
Individual stars in the ICL, such as intracluster planetary nebulae (IPNe), supernovae (ISNe), and red giants (IRGs), have been used to constrain the kinematics of the ICL as well as the total luminosity of intracluster stars \citep{Feldmeier1998a, Durrell2002a, Arnaboldi2004a, Feldmeier2004b, Gerhard2005,  McGee2010a, Sand2011a}. 
Deep observations of Coma, Virgo, and other intermediate redshift clusters show distinct tidal tails and bridges, suggesting that the ICL grows during dynamical exchanges and can be used as a measure of the evolutionary stage of the cluster \citep{Feldmeier2004a, Gerhard2005, Gonzalez2005, Mihos2005, KrickI, KrickII, Rudick2011}.  

Because ICL formation is closely linked to the processes of cluster assembly, its study provides a means by which to understand the processes involved in the evolution of galaxies in clusters. 
Four scenarios have been put forward as possible mechanisms by which stars are added to the the ICL: (1) the disruption of dwarf galaxies as they fall into the cluster potential, (2) the tidal stripping of the outskirts of \Lstar\ galaxies, (3) violent relaxation after major mergers between galaxies, including the central BCG, and (4) in situ star formation. 
Existing observational constraints on the envisioned formation scenarios are limited and theoretical models have yet to reach a consensus on which mechanism dominates the ICL formation process. 

The analytic models of \cite{purcell07} and simulations of  \cite{murante07} and \cite{conroy07} posit that the majority of the ICL is built up either through mergers and/or interaction with the BCG or through the shredding of dwarf satellites. 
Tidal stripping contributes only a small percentage of the ICL in the hydrodynamical simulations of \cite{murante07} and \cite{sommer} whereas it is a more dominant mechanism in  \cite{Rudick2009} and \cite{Watson2013a}'s n-body simulations and \cite{Laporte2013a}'s dark matter only simulations. 
\emph{In situ} formation is less favoured as a means to produce intracluster stars. 
\cite{Melnick2012}'s observations of the ICL of an intermediate redshift cluster attribute only 1\% of the ICL mass to this population of stars despite the predictions of \cite{Puchwein2010} that suggest \til30\% of the intracluster stars are formed \emph{in situ}.

The study of the ICL is potentially significant in the context of resolving classic debates about the drivers of galaxy evolution in dense environments. 
Potentially important processes include galaxy-galaxy tidal interactions and mergers, particularly in in-falling groups and subclusters (\cite{Zabludoff1996a, Zabludoff1998a}, now called "pre-processing"), encounters with the cluster potential, galaxy harassment, and ram pressure stripping (e.g., \cite{Park2009a, Smith2010a, Wezgowiec2012a}). 
If the ICL is dominated by stars stripped from galaxies, it offers something unique: a signature of the history of galaxy tidal interactions.  
By measuring the properties of the ICL over a range of redshifts and cluster masses, it is possible to constrain the importance of galaxy interactions in cluster evolution. 

In this work we use the surface brightness, infrared colour, and metallicity profiles of four massive galaxy clusters at intermediate redshift to comment on the progenitor population of the ICL and the dominant formation mechanism of the ICL. 
In \S\ref{sec:sample} we describe the cluster sample used for this study and in \S\ref{sec:datareduc} we present the data reduction process, including treatment of the sky background. 
In \S\ref{sec:sbnc} we derive the ICL surface brightness and colour profiles.
We next present the metallicity gradients in \S\ref{sec:mets} and discuss characteristics of individual clusters in \S\ref{sec:indiv} before presenting the physical implications and conclusions of our results in \S\ref{sec:discuss} and \S\ref{sec:concl}. 
Throughout the paper we use a cosmology with H$_0$=71 km s$^{-1}$ Mpc$^{-1}$, $\Omega_{M}$=0.27, $\Lambda$=0.73. 

\section{The Sample}
\label{sec:sample}

Our sample consists of clusters from the \lclash\ (CLASH, \cite{Postman2012a}).
 An \hst\ Multi-Cycle Treasury program, the CLASH sample consists of 25 galaxy clusters imaged in 16 different filters with WFC3/UVIS, WFC3/IR and ACS/WFC from 2000 $\AA$ to 17,000 $\AA$. The 524 orbits of observations for this program began in Cycle 18, with the program scheduled for completion during Cycle 20.

The CLASH sample is drawn mainly from the Abell and MACS cluster catalogs \citep{abell58, abell89, ebeling07, Ebeling2010a} and contains clusters with masses of 5$\times10^{14}$ - 3$\times10^{15}$ \msun that range in redshift from $z=0.18$ to $z=0.90$ with a median redshift of $z=0.4$. 
We refer the reader to \cite{Postman2012a} for additional details. 
In this paper, we focus on a subsample of four of the first observed CLASH clusters. 
With this subsample (Table \ref{table:samp}) we will study the evolution of the ICL for the upper mass range of galaxy clusters.

While ICL analysis is not one of the main science goals of CLASH, the data set is particularly well-suited for the task. 
Intracluster light is intrinsically a low surface brightness component of galaxy clusters and thus benefits greatly from \hst's low sky background relative to ground-based observatories and instrument stability. 
We focus on the IR observations from CLASH because the stellar populations of the ICL are old and red and the spectral energy distributions are significantly redshifted. 
Utilizing infrared filters allows us to probe the ICL out to lower surface brightnesses and greater cluster radii than ever before, which enables us to view the ICL outside of the region dominated by the cluster BCG.

Because sky subtraction is a critical step in the data reduction process, one aspect of CLASH that is not ideal for ICL analysis is the small field of view of WFC3. To minimize contamination by the ICL in the background estimation, we exclude the inner 300 kpc of the cluster, which limits the amount of sky area available for background measurement. 
We develop a sky subtraction method that produces robust background estimations, despite the small field of view of \emph{HST}, which we describe below in \S\ref{sec:skysub}.

\begin{table*}
\caption{Cluster Properties.}
\begin{threeparttable}[b]
\begin{tabular}{c c c c c}        
\hline \hline
Cluster  & z$_{clus}$ & M$_{500c}$ (10$^{14}$ M$_\odot$) & M$_{200c}$ (10$^{14}$ M$_\odot$) & Source \\ [0.5ex]
\hline \hline	
MACS1206.2-0847   & 0.440 & $10.6\pm2.1$ 	& $15.9\pm3.6$ & \cite{Umetsu2014a} \\
MACS0329.7-0211   & 0.450 & $7.7\pm1.1$ 		& $10.0\pm1.5$ & \cite{Umetsu2014a}\\
MACS1149.6+2232  & 0.544 & $14.2\pm3.4$ 	& $25.4\pm5.2$ & \cite{Umetsu2014a}\\
MACS2129.4-0741   & 0.570 & $10.6\pm1.4$ 	&  --  & \cite{Mantz2010a}\\
\end{tabular}
\end{threeparttable}
\label{table:samp}
\end{table*}

\section{Reduction and Analysis}
\label{sec:datareduc}
Using the standard \texttt{calwf3} package with reference darks, biases, and flats, we performed basic calibration for the data reduction. 
The first step of processing the data is performed as described in \cite{Schrabbac2010a}. 
We use MultiDrizzle, which uses time-dependent distortion solution from \cite{Anderson07} to produce distortion corrected exposures.  
Each exposure is then interactively aligned by cross-correlating the positions of compact sources and applying residual shifts and rotations to the exposures and fed back into MultiDrizzle.
The default values for MultiDrizzle's cosmic ray rejection are adjusted slightly to allow for PSF variation due to telescope breathing, which can cause central stellar pixels to be flagged as cosmic rays. 
However, we turn off sky subtraction to avoid removing the ICL and instead rely upon our own method to apply this vital step (see \S\ref{sec:skysub}). 
We use Scamp and SWaRP \citep{scamp, swarp} for a higher precision WCS alignment between epochs and different filters. 
A standard tangential projection is used in Scamp and LANCZ0S3 resampling is employed by SWaRP. 
A planar fit ensures minimal distortions from these two processes. 
Neglecting this step results in a $\sim$2-4 pixel misalignment. 
For our highest-redshift cluster this error would correspond to a misalignment of 2.6 kpc and introduce artificial features in the inner colour profile of the ICL.

\subsection{Key Considerations}
 \label{sec:issues}
Measurements of low surface brightness features like the ICL are intrinsically limited by background fluctuations. 
The two most dominant sources of uncertainties are flat fielding and background subtraction. 
Below we describe in detail our methodology for minimizing uncertainties and biases associated with each.

\subsubsection{Flat-Fielding}
\label{sec:flat}
Observations and analysis should be designed such that the error in photometry due to flat-fielding is negligible compared to the uncertainty due to sky subtraction.
Super sky flats, high precision flat fields produced with stacks of images first masked of all astronomical objects, are used as the pipeline flats in the reduction of our sample \citep{Pirzkal2011}. 
Across the whole WFC3 detector these flats produce an RMS error in the photometry of 0.7\% and a peak-to-peak uncertainty of 2.0$\pm$1.9\%. 
In the central 700 pix$^2$ of the CCD (excluding a border 1/8 the size of the detector) the RMS is 0.5\%, with a peak-to-peak uncertainty of 1.5$\pm$1.6\% \citep{WFC3}. 
For high surface brightness studies this level of systematic variation is insignificant; however, for the ICL this level of variation can introduce significant error. 

Two facts reduce the effects of low-frequency features in the flat field in our analysis. 
First, the CLASH field observations were taken at a variety of position angles. 
Field rotation distributes the observations over more pixels, thus averaging over large scale variation. 
Second, we focus on radially averaged quantities centered on the BCG (see \S \ref{sec:colorextract}) so within a bin the error in photometry  due to the field-fielding drops by a factor of 1$/\sqrt{N}$, where N is the number of uncorrelated pixels in each azimuthal bin. 

To verify that the uncertainty in photometry due to flat-fielding is not a dominant source of systematic uncertainty, we create preliminary `delta' flats for each filter by stacking WFC3/IR observations bracketing the observation dates of these CLASH clusters. 
We drizzle these delta flats in a manner identical to the drizzling of the science images (same number of input images, with the same rotation).  
From these drizzled flats we compute for each filter a radially averaged profile of the residual flatness variations using the same methodology as described in \S\ref{sec:sbnc}.  
We take the maximum amplitude of the radial flatness variation in the delta flats as the level of uncertainty introduced to the photometry due to variations in the flatness. Multiplying this variation by the measured background level for each epoch of data, we find the variation of the measured surface brightness due to flat fielding uncertainty. 
For all clusters in both F110W and F160W, the systematic uncertainty due to flat-fielding is sub-dominant to the uncertainty in the measured background level.

\subsubsection{Masking}
\label{sec:mask}
To accurately measure the ICL we must avoid contamination. 
The extended light from galaxies and stars smoothly merges into the light from the sky, so separating the ICL from these sources is a non-trivial challenge. 
The most robust approach is to mask out all other sources, making sure not to too aggressively mask sources and leave too few pixels for ICL measurements, or inadequately mask sources and leave contamination from stars or galaxies in with the ICL. 

\begin{figure*}
\centering
\subfloat{
\includegraphics[scale=0.33]{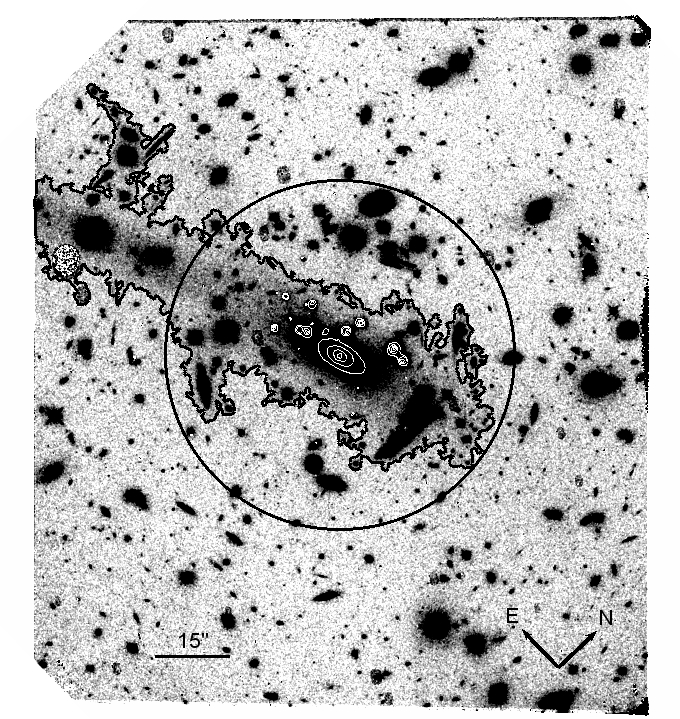}
}
\subfloat{
\includegraphics[scale=.33]{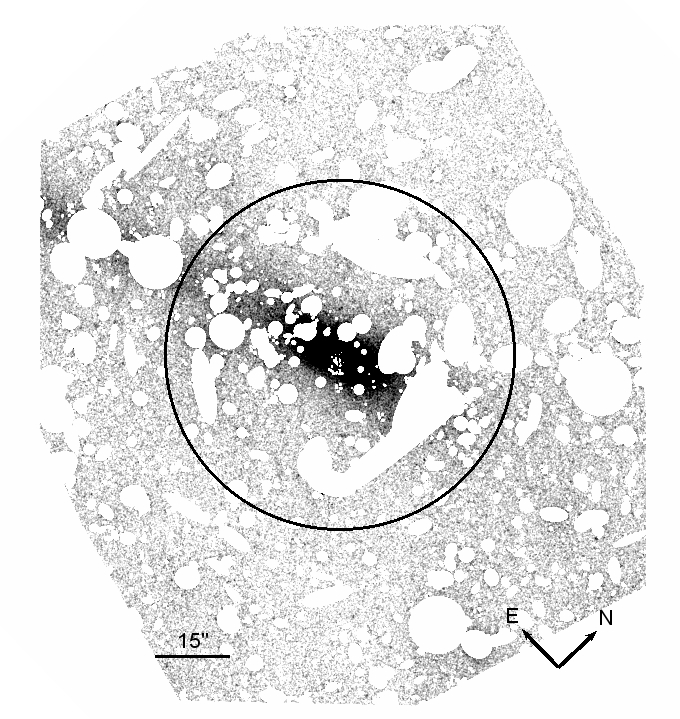}
}
\caption{Single WFC3 F160W image of \MACStwelve. The circle is centered on the cluster BCG and has a radius of 200 kpc. The ICL cannot be traced beyond this radius because it is too faint compared to the background signal. Black regions have surface brightness brighter than 25.7 \sbu\ and white contours are overlaid in the central region at 21-24 \sbu\ levels in steps of 1 \sbu. The right image is the same as at the left but after masking all galaxies and stars. Masked regions are excluded and the average of the sigma-clipped median of the remaining pixels in each background region is the background value for the image. The \til400 kpc low surface brightness bridge from the BCG to the north-east is a distinct feature, shown with a 27 \sbu\ level contour in black. Large-scale background structure is evident across all epochs of data for \MACStwelve; in the bottom corners of the above images, excess luminosity in the background is visible.}
\label{fig:m1206image}
\end{figure*}

\begin{figure*}
\centering
\subfloat{
\includegraphics[scale=0.33]{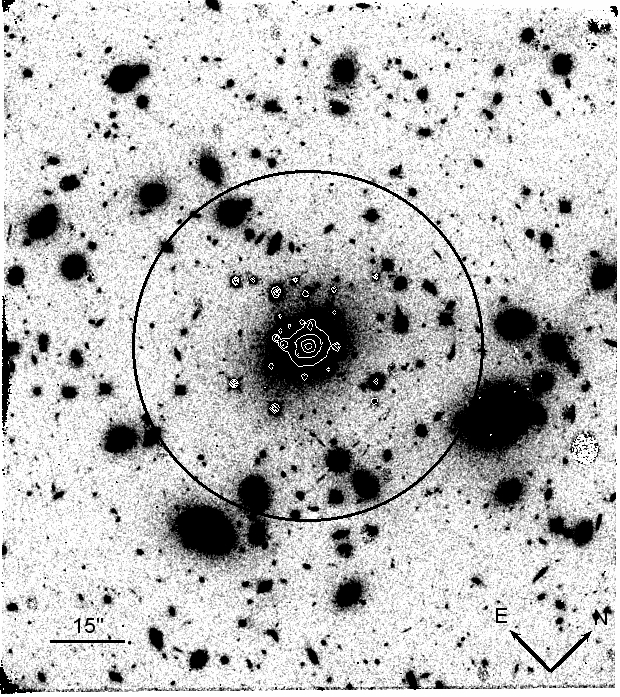}
}
\subfloat{
\includegraphics[scale=.33]{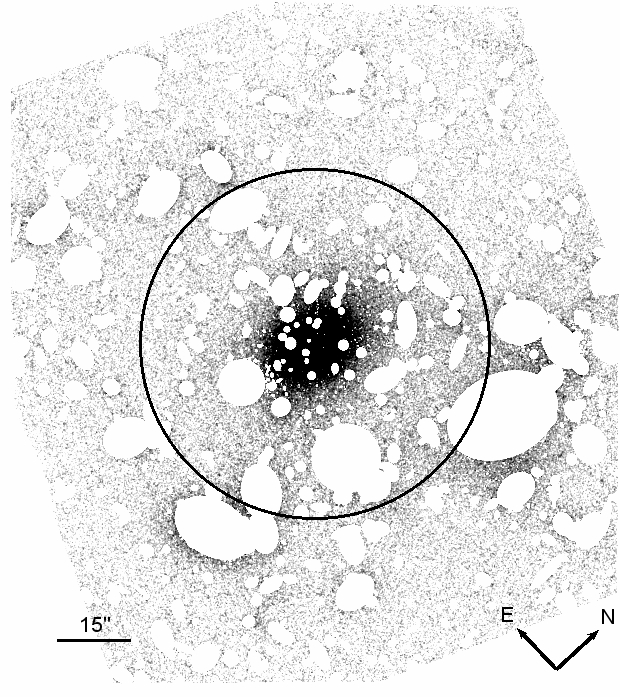}
}
\caption{Same as Figure \ref{fig:m1206image} but for \MACSohthree. Black regions have surface brightness brighter than 26.7 \sbu.}
\label{fig:m0329image}
\end{figure*}

\begin{figure*}
\centering
\subfloat{
\includegraphics[scale=0.33]{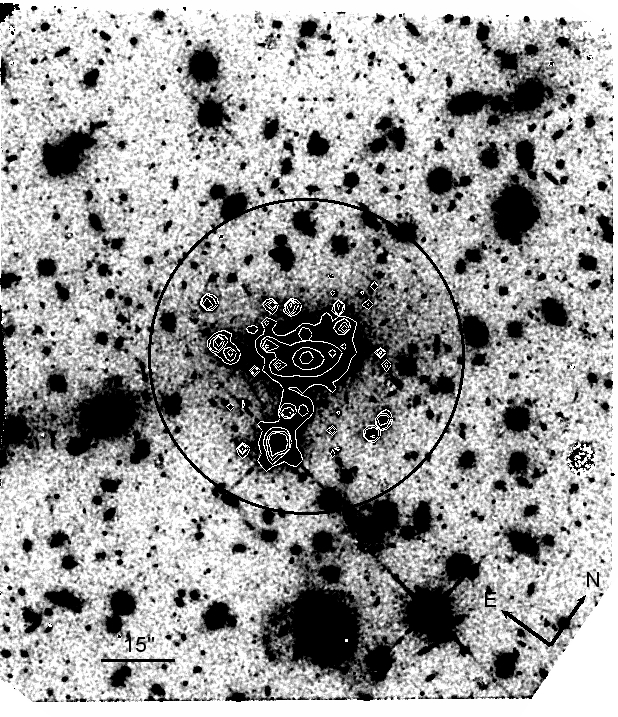}
}
\subfloat{
\includegraphics[scale=.33]{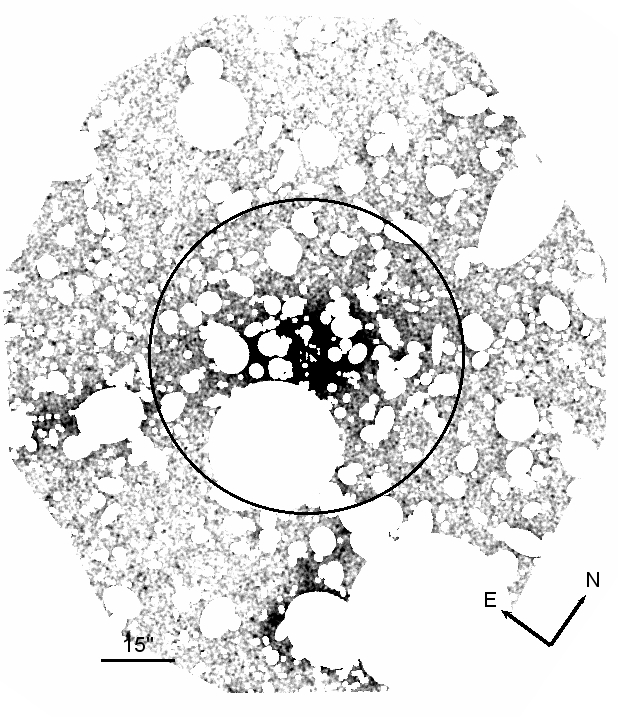}
}
\caption{Same as Figure \ref{fig:m1206image} but for \MACSeleven. Black regions have surface brightness brighter than 26.4 \sbu.}
\label{fig:m1149image}
\end{figure*}

\begin{figure*}
\centering
\subfloat{
\includegraphics[scale=0.33]{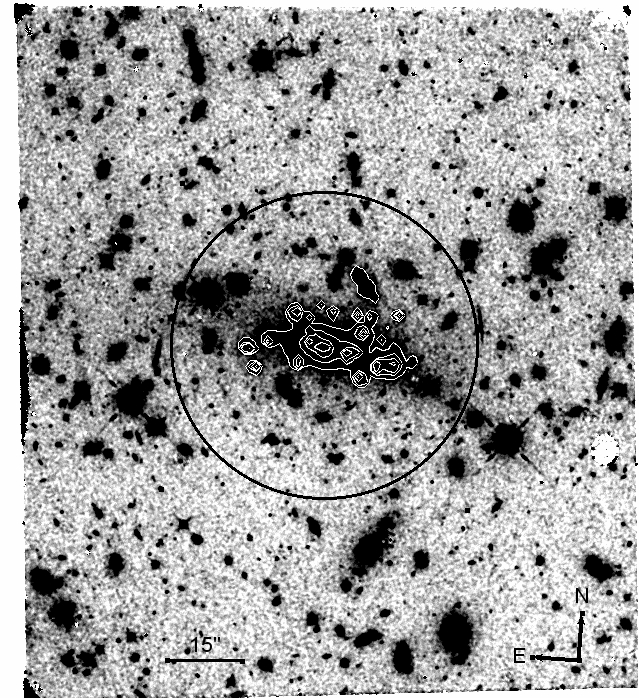}
}
\subfloat{
\includegraphics[scale=.33]{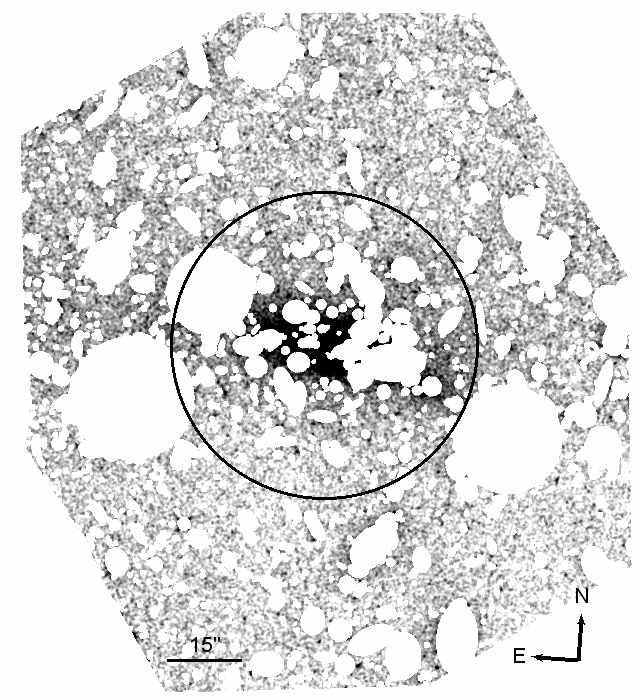}
}
\caption{Same as Figure \ref{fig:m1206image} but for \MACStwentyone. Black regions have surface brightness brighter than 25.7 \sbu.}
\label{fig:m2129image}
\end{figure*}

We approach this challenge via the following process. 
First, the sources are initially masked using ellipses extended out to 5 times the semi-major and semi-minor axes determined by Source Extractor \citep{SEx}.
A minimum detection area of five pixels and 2$\sigma$ detection limit per pixel are required. 
Near the BCG we do not use the automated Source Extractor masking. 
Instead, we manually mask any sources within 10\arcsec\ of the BCG by estimating the extent of the object by-eye and conservatively extending the mask radius to ensure no galaxy light remains in the final ICL sample. 
To ensure that the extended wings of stellar profiles are sufficiently masked, we enlarge stellar masks to 4\arcsec\ or to the extent of their diffraction spikes (\til10-12\arcsec\ for the brightest stars).
For even the brightest stars, this approach ensures that the light in the PSF wings is less than 2\%  of the sky level at 10\arcsec, which is below the uncertainty in the measured background level and thus does not contribute significantly to the measured ICL signal.
Each filter image is masked and the final cluster mask is produced by combining the individual filter masks, thus the same masks are used for both filters of a given cluster, enabling self-consistent colour measurements. 
This  method of masking is similar to those of \cite{Jee2010} and \cite{KrickI}.
See Figures \ref{fig:m1206image}-\ref{fig:m2129image} for pre- and post-masking examples.

\subsubsection{Background Subtraction}
\label{sec:skysub}
While \hst\ lacks the atmospheric emission that hinders ground-based observations, there is still background emission that must be considered. 
For pointings near the bright Earth limb, earthshine dominates over the contamination from zodiacal light and can produce a significant large-scale gradient in the background levels. 
It is also possible for \hst's optical system to scatter stray light from bright sources outside the field of view into images \citep{WFC3}. 
These effects cause low-level variation in the background among epochs. 
Because of this variation we cannot use the available parallel observations in F125W and F160W to estimate the background in our science images as they were not taken synchronously with the science observations. 
Instead, we determine the background for each epoch of data individually from the target image. 

We first fit a plane to the sky, subtract off the gradients in x and y, and leave the normalization to be fit more rigorously for each epoch. 
This plane is fit using the unmasked background region beyond 300 kpc from the BCG.
Next the image is sliced into 15 degree wedges, again excluding the inner 300 kpc and masked regions. 
For each wedge we determine the 3$\sigma$ clipped median of the unmasked pixels in that wedge using the standard deviation of all sky values within a wedge as the 1$\sigma$ error values. 
The weighted average of all slices of a given epoch is then the final value used for the background level of that epoch.

We find that the colour profiles produced by first subtracting the fitted background gradient are consistent with those produced without first subtracting the fitted planar background to $\ll1 \sigma$, suggesting that spatial gradients are not a significant driver of any measured large-scale gradients. 
We also look at the time-dependent nature of the measured background gradients and find them to be stable across the observations of a given cluster in a certain filter. 
We do this by differencing epochs of observations in a filter and fitting the residual background gradient using the same least-square technique. 
The residual gradient results in a total edge-to-edge difference in signal that is below the level of the measured background uncertainty.

The wedge method of determining the background has the advantage that any residual background gradient is averaged out by sampling the sky symmetrically. 
Because the sky is time-variable, we keep the data grouped by observation date in our analysis rather than combining after background subtraction. 
This choice has the additional benefit of allowing us to constrain the systematic errors between filters and observation dates.
See Table \ref{table:datadiv} for the measured sky values with uncertainties.

To test the robustness of our results, we use GALFIT \citep{GALFIT} to model the background. We first mask out the BCG+ICL component to 30'' on a masked, smoothed image, allow GALFIT to model the background with a planar model, and then renormalize the background values to have a median value equal to that found from the manual method described above. This alternate treatment yields results that are consistent at $<$1$\sigma$ to results from the manual method of sky subtraction. To use as few steps as possible for the reduction process, we use the manual method of sky subtraction only.

\begin{figure*}
\centering
\noindent\begin{minipage}{\linewidth}
\includegraphics[width=0.25\linewidth, height= 0.25\linewidth]{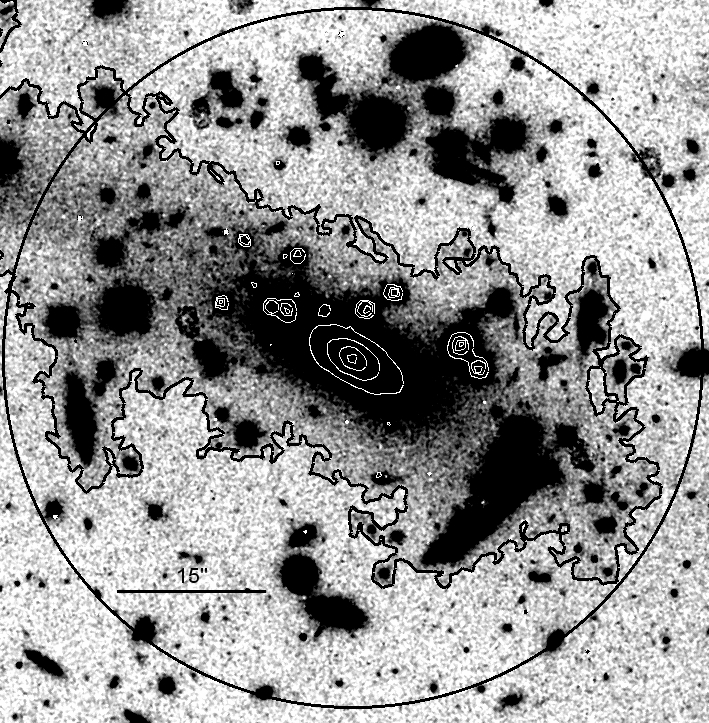}
\includegraphics[width=0.25\linewidth, height= 0.25\linewidth]{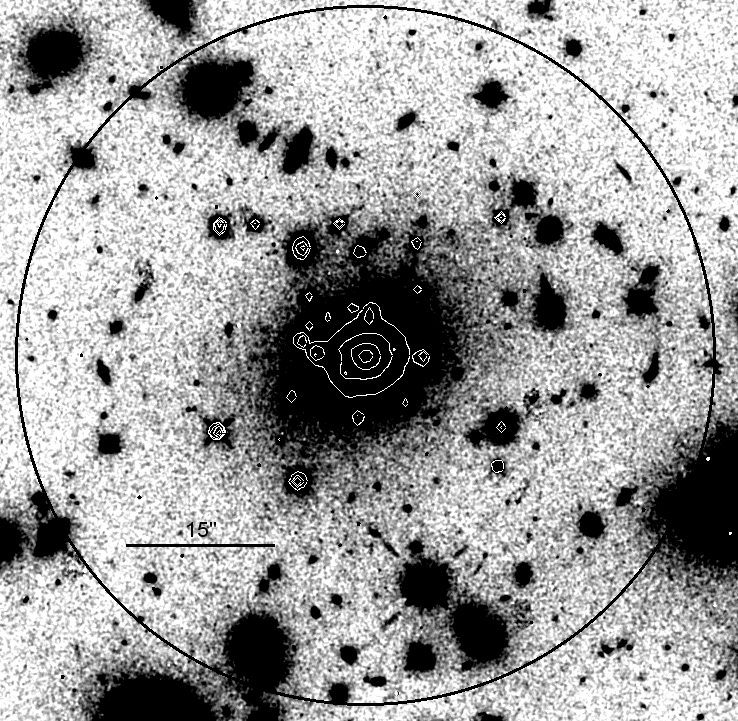}
\includegraphics[width=0.25\linewidth, height= 0.25\linewidth]{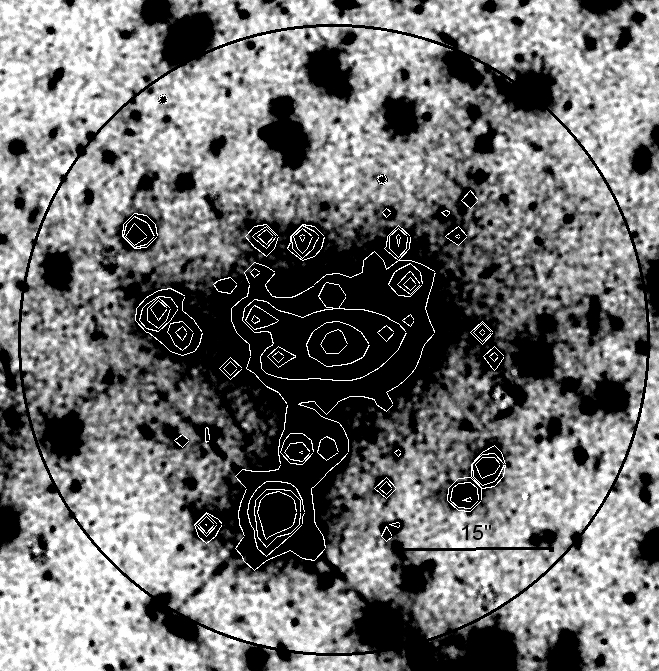}
\includegraphics[width=0.25\linewidth, height= 0.25\linewidth]{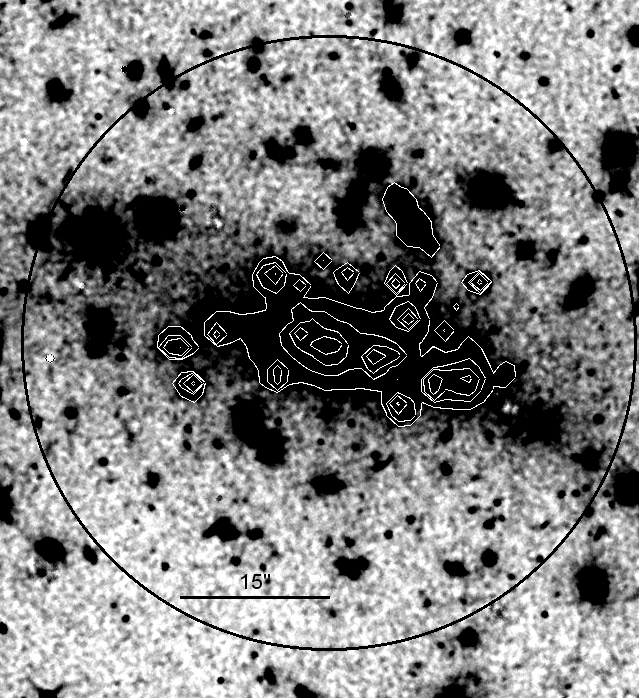}
\caption{The inner 200 kpc of \MACStwelve, \MACSohthree, \MACSeleven, and \MACStwentyone (left to right) in one epoch of F160W. White contours are at 21-24 \sbu\ in 1 \sbu\ increments and the black circle is at 200 kpc. Black regions are brighter than 25.7, 26.7, 26.4, and 25.7 \sbu for each cluster, left to right, respectively.}
\end{minipage}
\end{figure*}

\begin{table*}
\caption{Division of Observations.}
\centering
\begin{threeparttable}[b]
\scalebox{0.9}{
\begin{tabular}{c c c c c c c c c}
\hline\hline
Cluster & z & Filter & Date & PA$^1$ & Exposure & Sky & $\delta$Sky \\
& & & & [deg] & [sec] & \sbu & \sbu \\[0.5ex]
\hline
\MACStwelve  & 0.44 &  F160W  &  4/3/11  & -44.33 & 1005 & 23.11 & 28.65 \\
  &    &    &  6/21/11  & -44.33 & 1508 & 22.85 & 28.85\\
  &    &    &  5/2/11  & -15.33 & 1005 & 23.27 & 28.87 \\
  &    &    &  7/5/11  & -20.33 & 1508 & 22.69 & 28.62 \\
  &    &  F110W  &  4/3/11  & -44.33 & 1508 & 23.35 & 29.06 \\
  &    &    &  6/21/11  & -44.33 & 1005 & 23.09 & 29.37 \\
 \\
\MACSohthree  & 0.45 &  F160W  &  8/18/11  & 133.67 & 1005 & 23.09 & 29.66 \\
  &    &    &  10/16/11  & 133.67 & 1508 & 23.37 & 29.35 \\
  &    &    &  9/20/11  & 155.67 & 1005 & 23.32 & 29.37\\
  &    &    &  11/1/11  & 155.67 & 1508 & 23.23 & 29.15\\
  &    &  F110W  &  8/18/11  & 135.67 & 1508 & 23.27 & 30.13 \\
  &    &    &  10/16/11  & 135.67 & 502 & 23.50 & 30.45\\
  \\
\MACSeleven  & 0.544 &  F160W  &  1/16/11  & 124.67 & 1005 & 21.79 & 27.58 \\
  &    &    &  2/27/11  & 124.67 & 1508 & 21.79 & 27.71\\
  &    &    &  12/4/10  & 156.67 & 1005 & 21.46 & 27.62\\
  &    &    &  3/9/11  & 96.67 & 1508 & 21.79 & 27.67\\
  &    &  F110W  &  1/16/11  & 124.67 & 1508 & 21.98 & 28.52\\
  &    &    &  2/27/11  & 124.67 & 402 & 21.81 & 27.96\\
 \\
\MACStwentyone  & 0.57 &  F160W  &  5/15/11  & 94.67 & 1005 & 21.38 & 28.45 \\
  &    &    &  6/25/11  & 123.67 & 1205 & 21.81 & 28.37\\
  &    &    &  6/3/11  & 94.67 & 1408 & 21.65 & 28.38\\
  &    &    &  7/20/11  & 94.67 & 1408 & 21.76 & 28.50\\
  &    &  F110W  &  5/15/11  & 94.67 & 1408 & 21.50 & 28.52\\
  &    &    &  7/20/11  & 94.67 & 502 & 21.94 & 28.70\\
\hline
\end{tabular}}
\begin{tablenotes}[b]
	\item $^1$ Position angle from the reference aperture center.
\end{tablenotes}
\end{threeparttable}
\label{table:datadiv}
\end{table*}

\section{Surface Brightness and Colour Profiles}
\label{sec:sbnc}

\subsection{Surface Brightness Profiles}
\label{sec:sbu}

\begin{figure}
	\includegraphics[scale=0.6]{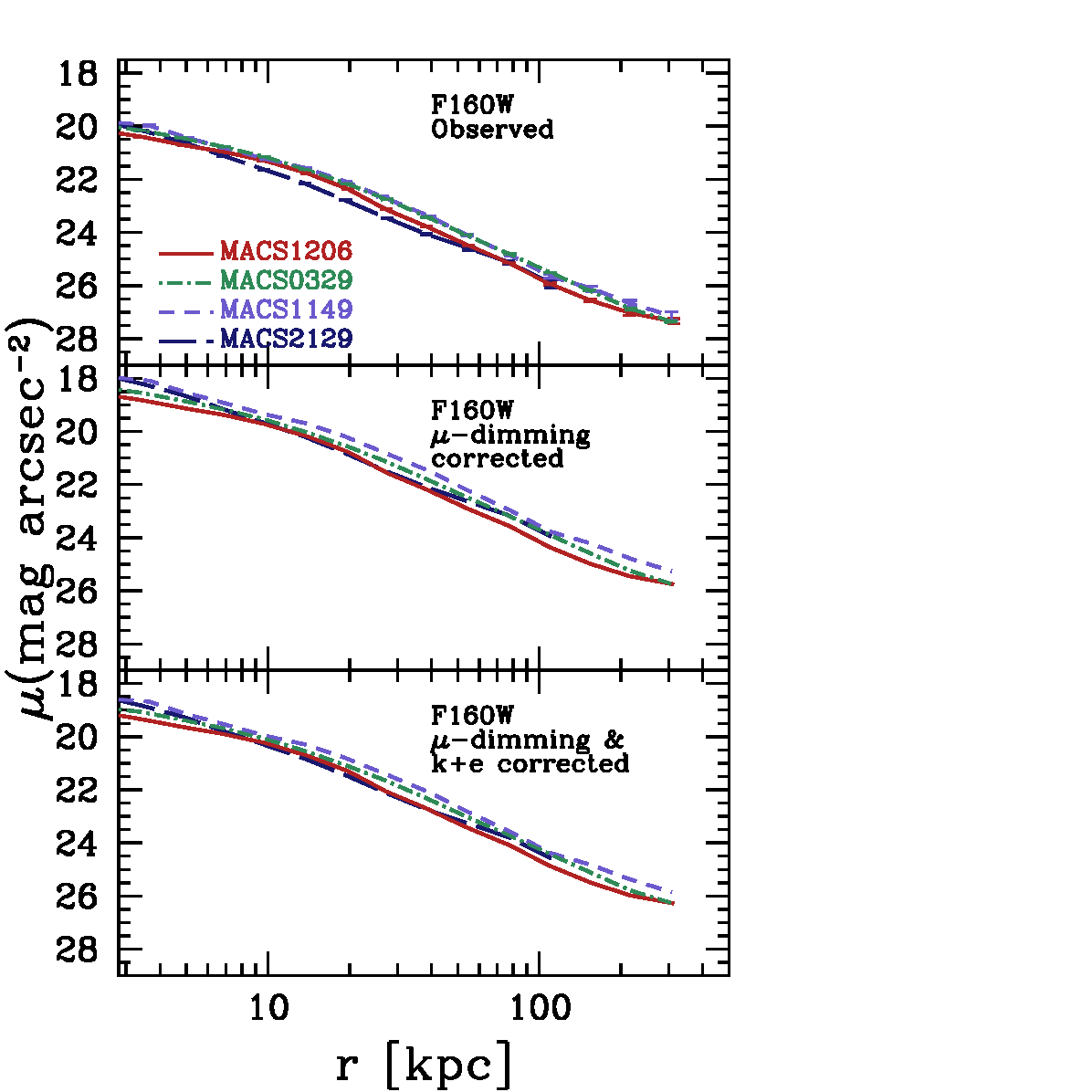}
\caption{F160W surface brightness profiles of BCG+ICL as a function of radius in dlog(r[kpc])=0.15 bins, out to the point where the error in \red\ surface brightness for each cluster is $<$0.2 \sbu and r$<$300 kpc.
The top panel shows the observed profiles with no passband or evolution correction. The middle figure takes the observed profiles and corrects for cosmological dimming. Finally, the bottom panel accounts for evolution and passband (e+k) corrections, assuming a BC03 model, as well as cosmological dimming.}
\label{fig:f160sb}
\end{figure}

Because the flat-fielding precision precludes a 2-D analysis of the surface brightness distribution at large radii, we produce radially averaged surface brightness profiles in \blue\ and \red\ for each cluster, maintaining a roughly uniform signal to noise by measuring the flux in logarithmically spaced annuli.
We take the median value of radii for unmasked pixels in a given bin as the bin location.
This can lead to slight differences (\til $\Delta$log(r[kpc])=0.007 in the locations of the bins between different clusters.
Any masked pixels are ignored in flux calculations and are not replaced. 
In the top panel of Figure \ref{fig:f160sb}, we present the raw observed F160W surface brightness profiles, terminating them when the error in the surface brightness in F160W is $>$0.2 \sbu or r$<$300 kpc.

Extending the background apertures inward to 300 kpc while simultaneously measuring the surface brightness profiles out to 300 kpc has the potential to result in over subtraction of the profile.  
However, we find that the impact of including the region at 300 kpc in the background annulus is negligible.
This is mainly because most of the weight in determining the sky level comes from larger radii, as the number of pixels contributing to the background estimate goes as r$^2$.
We have compared those background values as measured from 400 kpc outward to those from 300 kpc outward and find that they are the same within 1$\sigma$.

\begin{figure}
	\centering
	\includegraphics[scale=0.6]{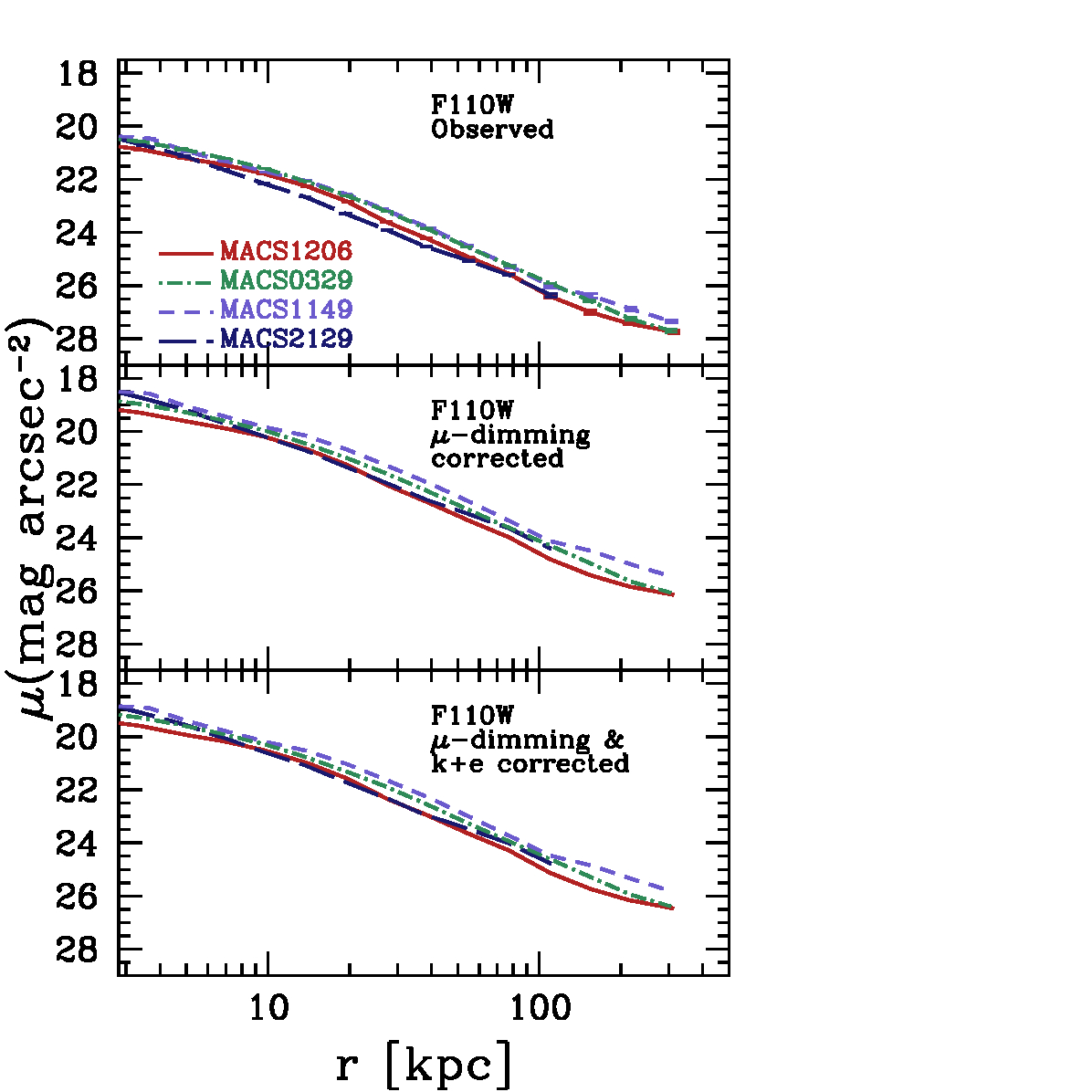}
\caption{The same as in Figure \ref{fig:f160sb}, but for \blue.}
\label{fig:f110sb}
\end{figure}

In the central panel of Figure \ref{fig:f160sb} we correct for cosmological dimming, enabling a more direct comparison of the profiles.
Finally, in the bottom panel we correct the observed surface brightness profile not only for cosmological dimming but also for evolution and passband (e+k) shifts. 
We use EZGAL \citep{ezgal} to produce e+k corrections for a \cite{BC03} (hereafter BC03) model with solar metallicity at a formation redshift of three (\zform=3) and Chabrier IMF; see Table \ref{table:metmodels} for exact parameters selected for each model. 
Using the evolution and passband corrections from the other available models has negligible effect on the final surface brightness profiles.
Figure \ref{fig:f110sb} is the same as Figure \ref{fig:f160sb} but for the \blue\ filter.

\begin{figure}
	\includegraphics[width=\columnwidth]{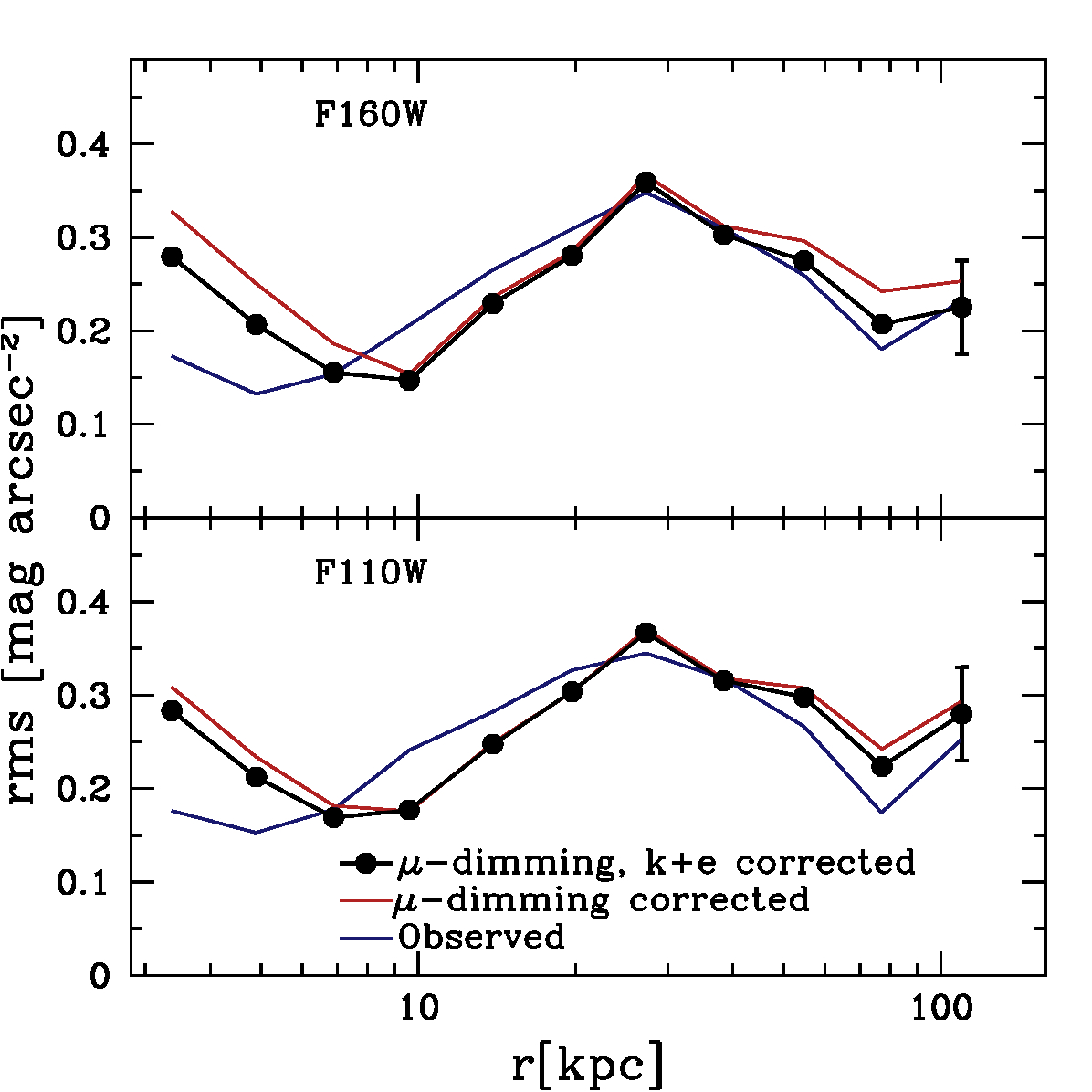}
\caption{The RMS scatter between the surface brightness profiles of all four clusters in F16W (top) and F110W (bottom) for each successive surface brightness correction. The scatter between observed profiles are traced by the blue line, the scatter between $\mu$-dimming corrected profiles by the red line, and completely corrected ($\mu$-dimming \& k+e corrected) in circles with a connecting black line. For 10$<$r$<$110 kpc the scatter remains between 0.20-0.37 \sbu in both filters, with a mean of 0.27 (0.29) \sbu\ in F160W (F110W). The error bar on the largest radius point reflects the estimated uncertainty in the plotted RMS values.} 
\label{fig:rmsboth}
\end{figure}

In Figure \ref{fig:rmsboth} we present the RMS scatter of the surface brightness profiles in bins of dlog(r[kpc])=0.15 in F160W (top) and F110W (bottom) for each successive correction (observed profiles in blue, $\mu$-dimming corrected in red, and $\mu$-dimming, passband, and evolution corrected as circles in black).
At r \ga 10 kpc, where the ICL is expected to dominate the surface brightness profiles, the scatter remains between 0.20-0.37 \sbu\ in both filters, with a mean value of 0.27 (0.29) \sbu\ in F160W (F110W).
A single representative error bar for the uncertainty in the RMS is included on the right-most point for both filters.
To factor out the differing masses of each cluster we scale the surface brightness profiles of all clusters to a uniform mass of 1$\times$10$^{15}$\Msun\ and find that the total RMS scatter in the profiles does not decrease.
This suggests that other factors, such as the specifics of an individual cluster's assembly history and ICL content, affect the scatter of observed surface brightness profiles.

\subsection{Colour Profile Extraction}
\label{sec:colorextract}

\begin{figure}
\centering
\includegraphics[width=0.5\textwidth]{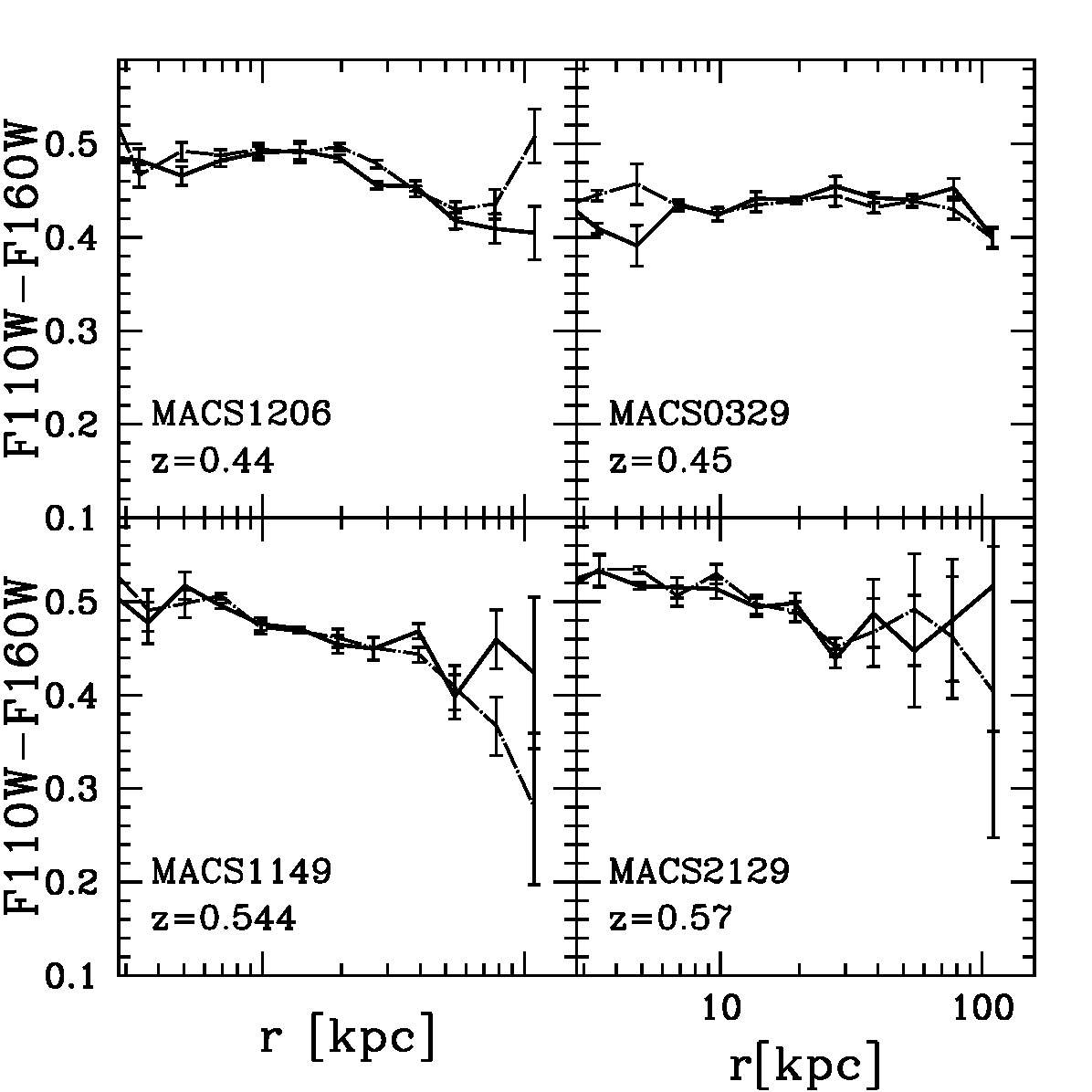}
\caption{F110W-F160W observed colour profiles, with no passband or evolution correction applied, for all four clusters in order of increasing redshift, top to bottom. We produce 2 averaged colour profiles, one for each blue filter epoch of data (solid and dashed line, respectively). All radii with $\mu_{F160W}<26$ \sbu\ are plotted with error bars. ICL measurements are sensitive to background subtraction and thus our largest source of systematic uncertainty is from sky brightness variation among epochs of data. Differences between filters and observation dates drive the larger error bars and differences in colour seen in the colour profiles at large radii.}
\label{fig:allcolor}
\end{figure}

We present the BCG+ICL colour profiles that we produce by subtracting the surface profiles of each of the two filters bin by bin  in Figure \ref{fig:allcolor}. 
The same masks are used for each filter, thus any resulting features in the colour profile are not due to a lack of symmetry in masking. 
We plot all data to F160W=26 \sbu, which effectively limits our profiles to r$<$110 kpc.
Beyond this point the colour profiles become too noisy for use in robustly determining ICL properties.

Due to the sensitivity of ICL measurements to background subtraction and the time variability of the sky, our largest source of systematic uncertainty is from sky brightness variation among epochs. 
The background level can vary by as much as 53\% in F160W and 40\% in F110W between observation epochs of a cluster in a given filter. 
(See Table \ref{table:variation} for details of sky level variation per cluster per filter and Table \ref{table:datadiv} for background surface brightness levels with uncertainties for each epoch of data.)

Large-scale background structure is evident across all epochs of data for both \MACSeleven\ and \MACStwelve, which is also in a region of high zodiacal light.
Additionally, there is known Galactic cirrus north-east of the cluster centre of \MACStwentyone \citep{Postman2012a}.
The increased uncertainty in the colour profiles of these clusters at radii $>50-100$ kpc is a reflection of the presence of this structure.
While both \MACStwelve\ and \MACSeleven\ show large-scale gradients in the background level, these gradients are stable across all epochs of data and thus do not alter the observed colour gradient beyond any effect that remains after symmetrically sampling the sky to determine each epoch's background level (as described in \S\ref{sec:skysub} above).

To understand the effect of systematic errors between different filters and observation dates we analyse each epoch of data individually and produce multiple colour profiles per cluster. We then average together all the profiles produced with a single F110W  epoch of data, e.g.
\begin{eqnarray}
<\blue-\red>_i=\displaystyle \sum_j^{N_j} \frac{(\blue_i-\red_j)}{N} \nonumber \\
=\blue_i-\displaystyle \sum_j^N\frac{\red_j}{N} 
\end{eqnarray}
We use the standard deviation of the mean as an estimate of the uncertainty because our systematics far outweigh our statistical errors.
 A tabular version of the final F110W-F160W profile with 1$\sigma$ error values in dlog(r[kpc])=0.15 binning of each cluster is presented in Table \ref{table:color}.

\begin{table}
\caption{Sky variation between epochs.}
\begin{threeparttable}[b]
\begin{tabular}{c c c c}        
\hline \hline
Cluster  & Filter & Sky variation & $\delta$Sky \\
		 &		  & \%			  & \sbu		\\ [0.5ex]
\hline
\MACStwelve 	& F160W & 53.1 & 27.98 \\
 				& F110W & 23.9 & 28.81 \\
 \MACSohthree 	& F160W & 26.0 & 28.60 \\
			 	& F110W & 21.6 & 29.89 \\
\MACSeleven 	& F160W & 32.3 & 26.89 \\
				& F110W & 15.2 & 27.73 \\
\MACStwentyone 	& F160W & 40.6 & 27.67 \\
				& F110W & 39.7 & 28.23 \\
\hline 
 \end{tabular}
\end{threeparttable}
\label{table:variation}
\end{table}

\begin{figure}
\centering
\includegraphics[width=0.5\textwidth]{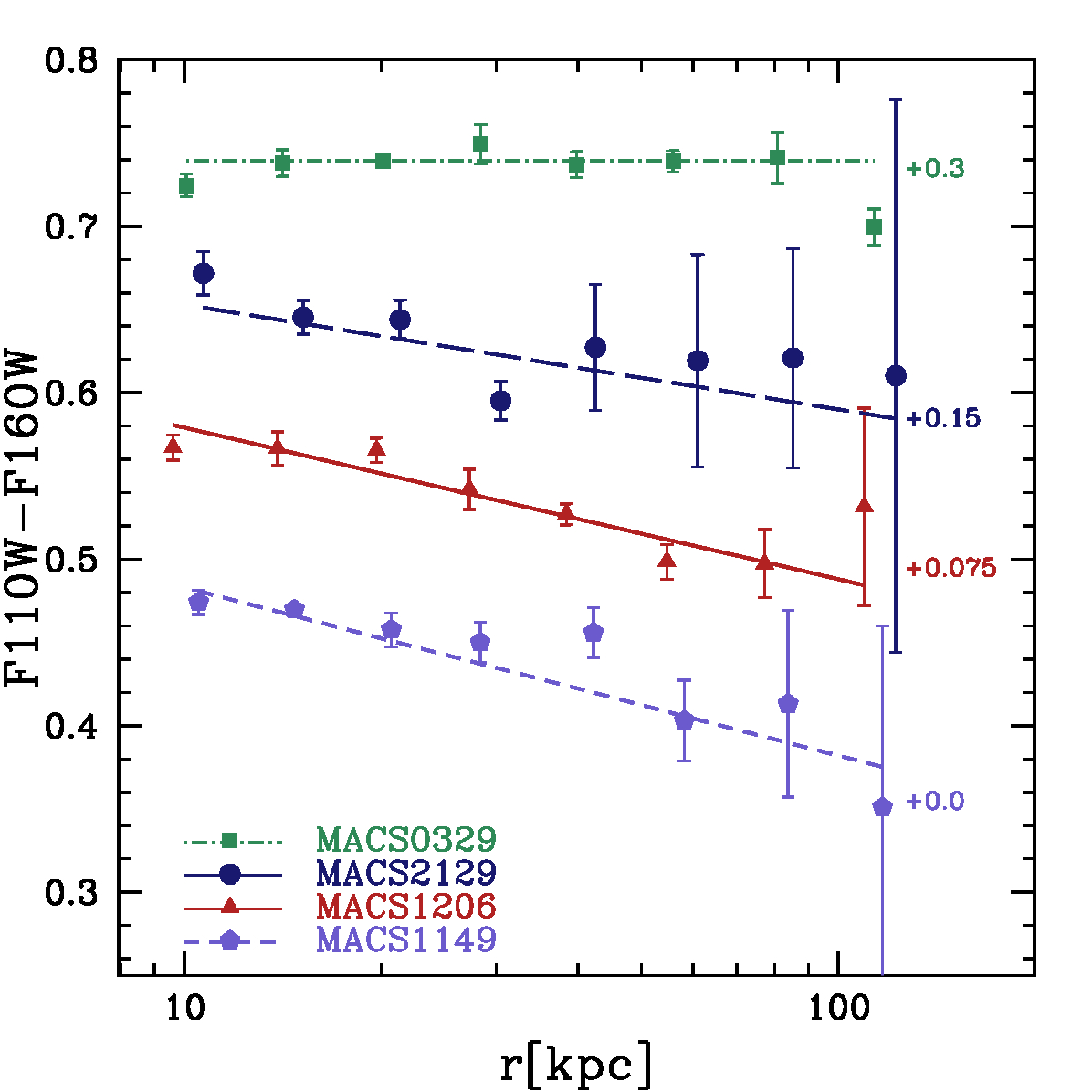}
\caption{The observed F110W-F160W averaged colour profile of each cluster, arranged by increasing slope, top to bottom. The large points are the colours of each cluster in dlog(r[kpc])=0.15 with error bars representing the standard deviation of the mean colour. Clusters are offset in colour by the amount given at the right side of the figure and the best-fit linear equations resulting from fitting the dlog(r[kpc])=0.01 profiles for each cluster are overlaid. All clusters except \MACSohthree\ trend toward significantly bluer colours at higher radius with a colour gradient \colorgrad\ between -0.063 and -0.101 mag log(kpc)$^{-1}$ (see Table \ref{table:fact}).}
\label{fig:allover}
\end{figure}

We investigate the ICL colour gradient by creating a single, averaged profile from all the individual colour profiles of each cluster. 
To maintain a uniform radial range for all clusters and produce a robust analysis, we fit only those radii for all clusters where the F160W surface brightness is < 26 \sbu, which restricts our fits to r$<$110 kpc.
Further, we do not fit inside 10 kpc, where the BCG dominates the surface brightness profiles.
Using a weighted linear fit to constrain each profile for 10 $\leq$ r [kpc] $\leq$ 110, we find that the measured colour gradients of three of the four clusters are negative at the $\ge$3$\sigma$ level (Table \ref{table:fact}.). 
To derive these values we use radial profiles with dlog(r[kpc])=0.01 in order to avoid ambiguity in the discretization of the bins which may drive the fitted values.
In Figure \ref{fig:allover} we plot the observed colour gradients in order of increasing slope, top to bottom, and overlay the best fit, offset in colour for clarity. 
The large points are the colours of each cluster in dlog(r[kpc])=0.15, and the fitted lines are the results from fitting the dlog(r[kpc])=0.01 profiles.
All clusters, with the exception of \MACSohthree, trend toward significantly bluer colour at larger cluster radii. 
To test the time stability of the colour gradients we also fit the colour profiles resulting from averaging together the F110W-F160W profiles generated with each epoch of F110W and compare the resulting slope to that produced from the average of the colour profiles from all of epochs.
The resulting slopes from all fits are consistent for all clusters, indicating that the reported colour gradients are not time dependent.

\begin{table*}
\caption{Colour Profiles}
\centering
\begin{threeparttable}[b]
\scalebox{0.9}{
\begin{tabular}{c c c c c c c c}
\hline\hline
\multicolumn{2}{c}{\MACStwelve} & \multicolumn{2}{c}{\MACSohthree} & \multicolumn{2}{c}{\MACSeleven} & \multicolumn{2}{c}{\MACStwentyone} \\
log(r[kpc]) & F110W-F160W & log(r[kpc] & F110W-F160W & log(r[kpc] & F110W-F160W & log(r[kpc] & F110W-F160W \\
\hline
0.08	&	0.489	$\pm$	0.031	&	0.10	&	0.496	$\pm$	0.018	&	0.11	&	0.497	$\pm$	0.038	&	0.08	&	0.468	$\pm$	0.026	\\
0.24	&	0.484	$\pm$	0.016	&	0.24	&	0.442	$\pm$	0.059	&	0.21	&	0.494	$\pm$	0.020	&	0.23	&	0.469	$\pm$	0.024	\\
0.39	&	0.523	$\pm$	0.036	&	0.37	&	0.438	$\pm$	0.016	&	0.39	&	0.531	$\pm$	0.027	&	0.38	&	0.513	$\pm$	0.028	\\
0.53	&	0.474	$\pm$	0.015	&	0.53	&	0.427	$\pm$	0.018	&	0.56	&	0.484	$\pm$	0.023	&	0.54	&	0.534	$\pm$	0.017	\\
0.69	&	0.479	$\pm$	0.016	&	0.68	&	0.424	$\pm$	0.040	&	0.70	&	0.507	$\pm$	0.018	&	0.69	&	0.525	$\pm$	0.009	\\
0.84	&	0.485	$\pm$	0.007	&	0.84	&	0.434	$\pm$	0.005	&	0.84	&	0.501	$\pm$	0.006	&	0.83	&	0.510	$\pm$	0.012	\\
0.98	&	0.492	$\pm$	0.008	&	0.99	&	0.425	$\pm$	0.007	&	0.99	&	0.474	$\pm$	0.007	&	0.98	&	0.522	$\pm$	0.013	\\
1.14	&	0.492	$\pm$	0.010	&	1.14	&	0.438	$\pm$	0.008	&	1.14	&	0.470	$\pm$	0.002	&	1.14	&	0.495	$\pm$	0.010	\\
1.29	&	0.491	$\pm$	0.007	&	1.29	&	0.439	$\pm$	0.003	&	1.29	&	0.458	$\pm$	0.010	&	1.29	&	0.494	$\pm$	0.012	\\
1.44	&	0.467	$\pm$	0.012	&	1.44	&	0.449	$\pm$	0.012	&	1.42	&	0.450	$\pm$	0.012	&	1.44	&	0.445	$\pm$	0.012	\\
1.58	&	0.452	$\pm$	0.006	&	1.59	&	0.437	$\pm$	0.008	&	1.60	&	0.456	$\pm$	0.015	&	1.58	&	0.477	$\pm$	0.038	\\
1.74	&	0.423	$\pm$	0.010	&	1.73	&	0.439	$\pm$	0.006	&	1.73	&	0.403	$\pm$	0.024	&	1.74	&	0.469	$\pm$	0.064	\\
1.89	&	0.422	$\pm$	0.020	&	1.89	&	0.441	$\pm$	0.015	&	1.89	&	0.413	$\pm$	0.056	&	1.89	&	0.471	$\pm$	0.066	\\
2.04	&	0.457	$\pm$	0.059	&	2.04	&	0.399	$\pm$	0.011	&	2.04	&	0.351	$\pm$	0.109	&	2.04	&	0.460	$\pm$	0.166	\\
\hline

\end{tabular}}
\end{threeparttable}
\label{table:color}
\end{table*}

\begin{figure}
\centering
\includegraphics[width=0.5\textwidth]{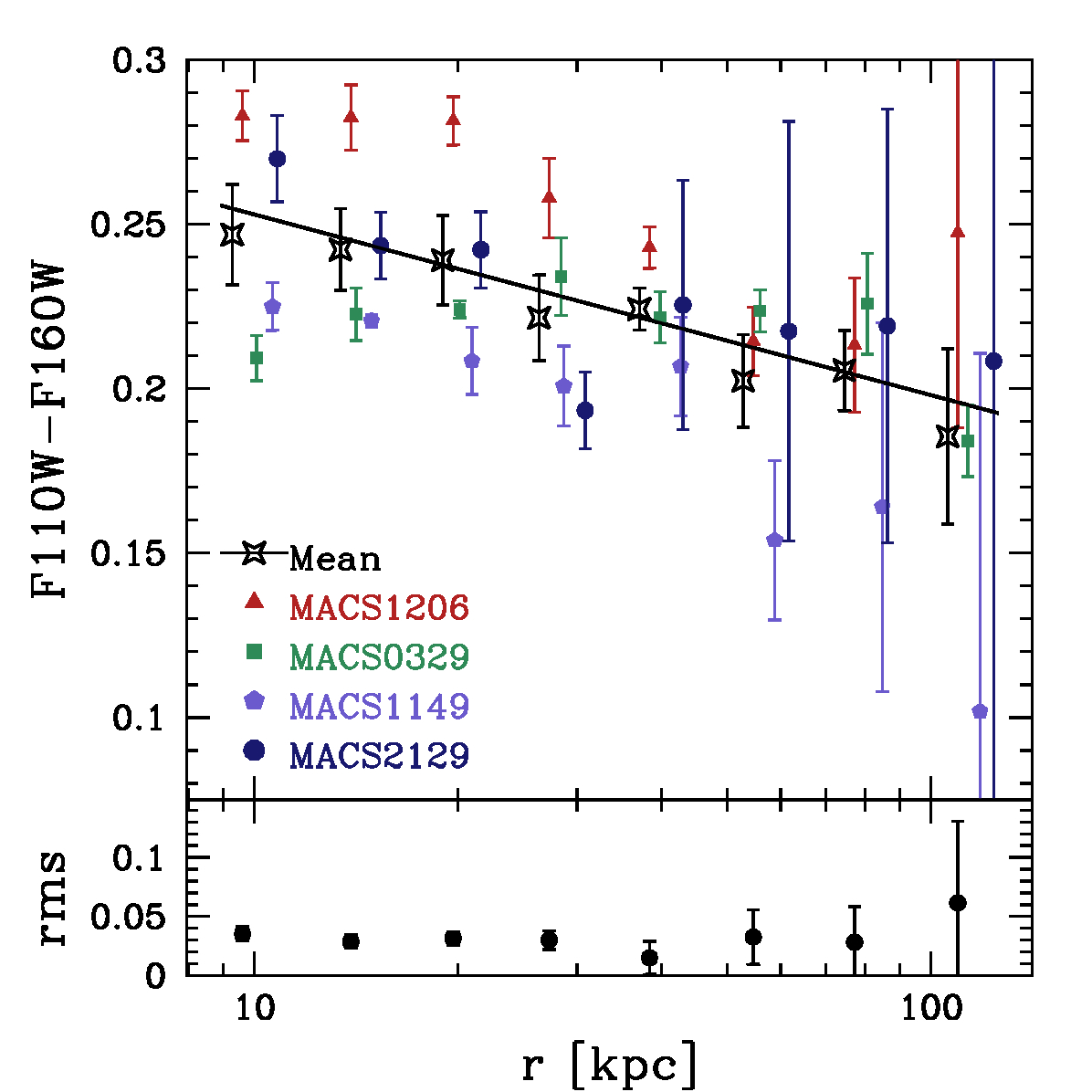}
\caption{The \red-\blue\ colour profiles after passband and evolution correction with a BC03 model for a solar metallicity stellar population formed at \zform=3. Each cluster is slightly offset in radius for clarity. The large black points are the mean in each dlog(r[kpc])=0.15 bin with the standard error in the mean as the 1$\sigma$ error bars. The black line represents the linear fit found for the mean colour vs. radius relationship. The RMS scatter between the colour profiles of the four clusters is shown in the lower panel, which is low and roughly uniform from 10$<$r$<$110 kpc.}
\label{fig:color_rms}
\end{figure}

We next apply passband and evolution corrections, derived from a BC03 model for a solar metallicity stellar population formed at \zform=3, to the observed \blue$-$\red\ colour profiles to bring them to their z=0 colours.  
The resulting profiles are shown in Figure \ref{fig:color_rms}, where we plot each cluster slightly offset in radius for clarity. 
The mean colour of all four clusters in each dlog(r[kpc])=0.15 bin is plotted in large black triangles.
We perform a linear fit to the mean colour profile, resulting in a colour gradient of \colorgrad=$-$0.055$\pm$0.010 \sbu\ log(kpc)$^{-1}$.

\section{Metallicity Gradients}
\label{sec:mets}

Evolutionary models \citep{Melnick2012,Rudick2006,sommer,Murante2004a,murante07,conroy07} and observations \citep{Feldmeier2004a,KrickI} suggest that the formation of the ICL is closely related to the dynamical history of the cluster, thus its study provides valuable insight into the cluster formation process. 
One way to distinguish between the many ICL and cluster assembly scenarios is to look at the metallicity of the ICL. 
The existence of a mass-metallicity relation \citep{Skillman1996a} in cluster galaxies implies that one can identify the characteristic mass of the progenitor population using the metallicity of the ICL.
Intracluster stellar metallicities have been spectroscopically measured in Virgo and Fornax from several types of intracluster populations including RGB stars, planetary nebulae, and supernovae \citep{Durrell2002a, Feldmeier1998a,Williams2007a, Coccato2011a}. 
In general, these studies find that intracluster stars are between 2-13 Gyr old and have sub-solar metallicities (-0.8$<$[Fe/H]$<$-0.2) \citep{Durrell2002a, Coccato2011a}. 
We use the colour gradients of the integrated stellar population presented in \S \ref{sec:colorextract} to constrain the metallicity profiles of the ICL, modulo uncertainty in the age of the stellar population.

\subsection{Stellar Population Models}
\label{sec:pops} 

To transform the observed colour gradients into metallicity profiles, we employ stellar population synthesis (SPS) modeling. 
We use five widely-used model sets, taking the models of \cite{BC03} as the fiducial to examine the effect of model choice on our results. 
These include an updated version of the BC03 model (often referred to as CB07) and the model set from \cite{M05} (hereafter M05). 
Both include detailed treatment of thermally pulsating asymptotic giant branch stars (TP-AGB), which can have a potentially significant effect on the infrared emission of intermediate age stellar populations. 
Additionally, we use Flexible Stellar Population Synthesis models from \cite{C09} and \cite{C10} (hereafter C09), which treat some SPS inputs, such as IMF and stellar evolution uncertainties, as free parameters. 
We also use the models from \cite{BaSTI} (hereafter BaSTI), which have a larger range of metallicities and $\alpha$-enhanced models. 
Table \ref{table:metmodels} lists available metallicities and IMFs for each model set. 
We primarily use a Chabrier IMF \citep{Chabrier2003} for the BC03 and CB07 model sets and Kroupa IMF  \citep{Kroupa2001} for the C09, M05, and BaSTI model sets. 
We employ a simple stellar population for our star formation history.
From these SPS models we derive passband and evolution corrections (see \S\ref{sec:sbnc}), as well as metallicities of the ICL given an observed colour (see \S\ref{sec:metal}). 

The clear blue-ward radial trend in the \blue--\red\ colour profiles seen in Figure \ref{fig:allover} can be interpreted as either an age-radius or a metallicity-radius relation, however there exist both theoretical and observational reasons to expect that the dominant factor is metallicity. 
Observationally, local studies generally indicate that the intracluster stellar population is an old stellar population ($>$10 Gyrs, \cite{Williams2007a, Coccato2011a, Melnick2012, Loubser2009}). 
Theoretically, models predict that the intracluster stars form from z\til2 \citep{Murante2004a} to z\til3 \citep{sommer, Puchwein2010}. 
Because of the general consensus that the ICL is an old stellar population we explore the interpretation that our observed colour gradients are due to changes in metallicity, noting the potential impact of the age-metallicity degeneracy.

\begin{table*}
\centering
\caption{Model Set Parameters. Adapted from \cite{ezgal}, Table 1. The primary IMF model used for each stellar population synthesis model in our analysis is bolded.}
\begin{tabular}{c c c c c c c}
\hline\hline
Model & Metallicity ($Z/Z_\odot$) & $\alpha$-enhanced & \# Metallicities & Salpeter & Chabrier & Kroupa \\ [0.5ex]
\hline
BC03 & 0.005-2.5 & No & 6 & Yes & \textbf{Yes} & No \\ 
CB07 & 0.005-2.5 & No & 6 & Yes & \textbf{Yes} & No \\ 
BaSTI & 0.005-2 & Yes & 10 & No  & No & \textbf{Yes} \\ 
C09 & 0.01-1.5 & No & 22 & Yes & \textbf{Yes} & Yes \\ 
M05 & 0.05-3.5 & No & 5 & Yes & No  & \textbf{Yes} \\ [0.5ex]
\hline
\end{tabular}
    \label{table:metmodels}
\end{table*}

\subsection{Metallicity Profiles}
\label{sec:metal}
We transform our observed colour gradients to metallicity profiles with each of the SPS models, using EZGAL \citep{ezgal}, which outputs observable parameters from input SPS models. 
For a given model set, cluster redshift, assumed formation redshift, and star formation history, the metallicity of a given stellar population has a unique corresponding colour. 
Over the range of metallicities available for each model we build up a metallicity-colour function for each model set. 
The ICL metallicity gradient of a cluster is then found by interpolating the observed colour in each radial bin to a corresponding model-derived metallicity. 

\begin{figure*}
\centering
\includegraphics[width=\textwidth]{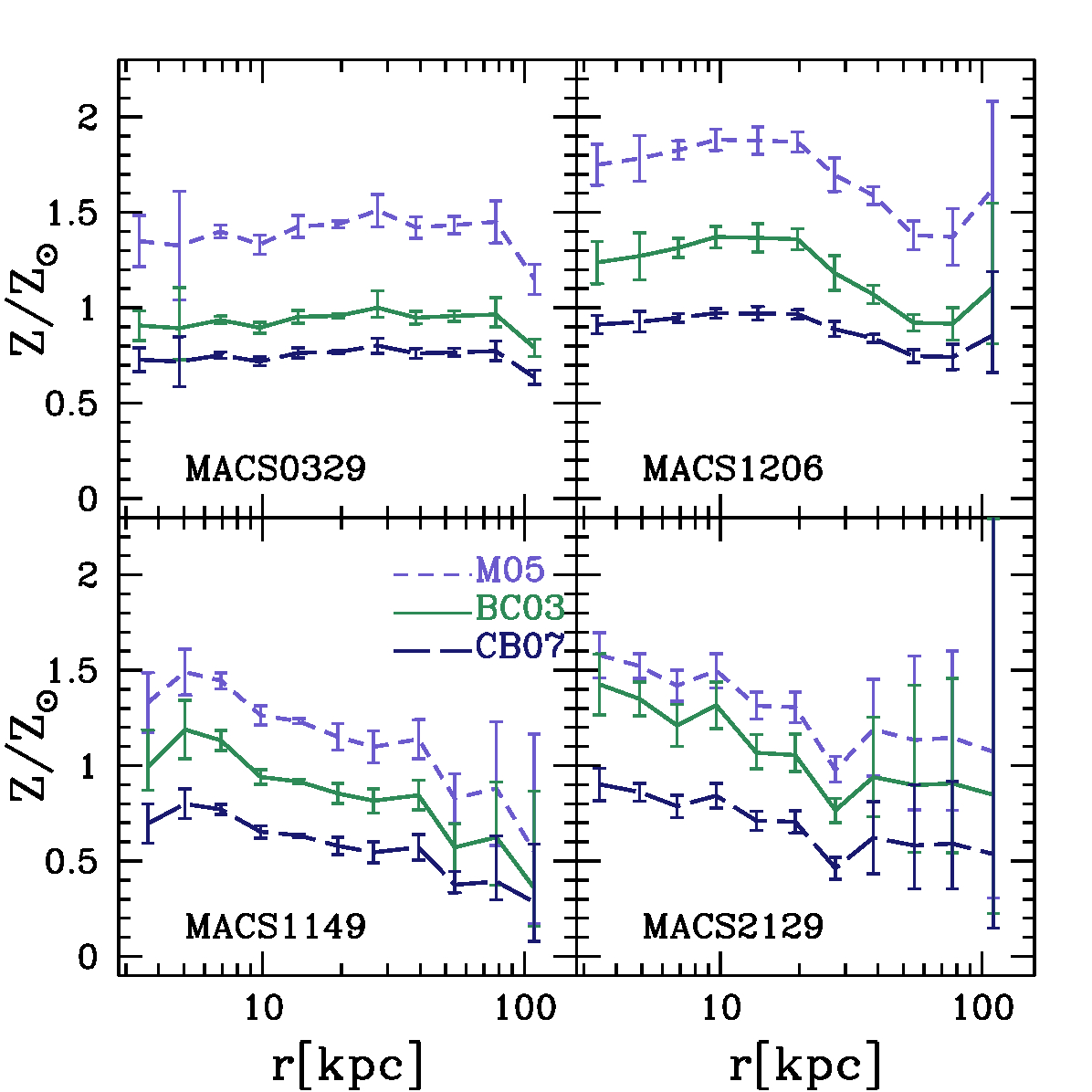}
\caption{SPS models are employed to transform the observed F110W-F160W dlog(r[kpc])=0.15 colour profile for each cluster to metallicity as a function of radius. Errors are those errors from the colour profiles in Figure \ref{fig:allover}, transformed to metallicity in the same fashion. In general, there is considerable spread between the metallicity profiles produced with the five stellar population models, though the basic shape of each profile remains the same. We show only BC03, CB07, and M05 for clarity.}  
\label{fig:met_panel}
\end{figure*}

The colour gradients in Figure \ref{fig:color_rms} directly translate into corresponding metallicity gradients under the assumption of a fixed age for the ICL.
Using a formation redshift of \zform=3 as a baseline, the resulting metallicity gradients are shown in Figure \ref{fig:met_panel}. 
We use the \blue$-$\red\ colour to interpolate to metallicity as \blue\ is less noisy than \green\ and \blue$-$\red\ provides greater wavelength leverage. 
Of the five models, we show only M05, BC03, and CB07 in Figure \ref{fig:met_panel}.
BC03 is our fiducial model and generally lies between the more extreme predictions of M05 and CB07. 
For clarity, we do not show the metallicity gradients for C09 and BaSTI, which fall between the models shown and exhibit the same general shape. 
The $\alpha$-enhanced BaSTI models produce gradients consistent with the zero $\alpha$-enhancement BaSTI models. 
We investigate the effects of IMF choice on the produced metallicity gradients and find little effect; the gradients produced with a Salpeter IMF are consistent with those produced using a Chabrier or Kroupa IMF.

\begin{figure}
\centering
\includegraphics[width=0.5\textwidth]{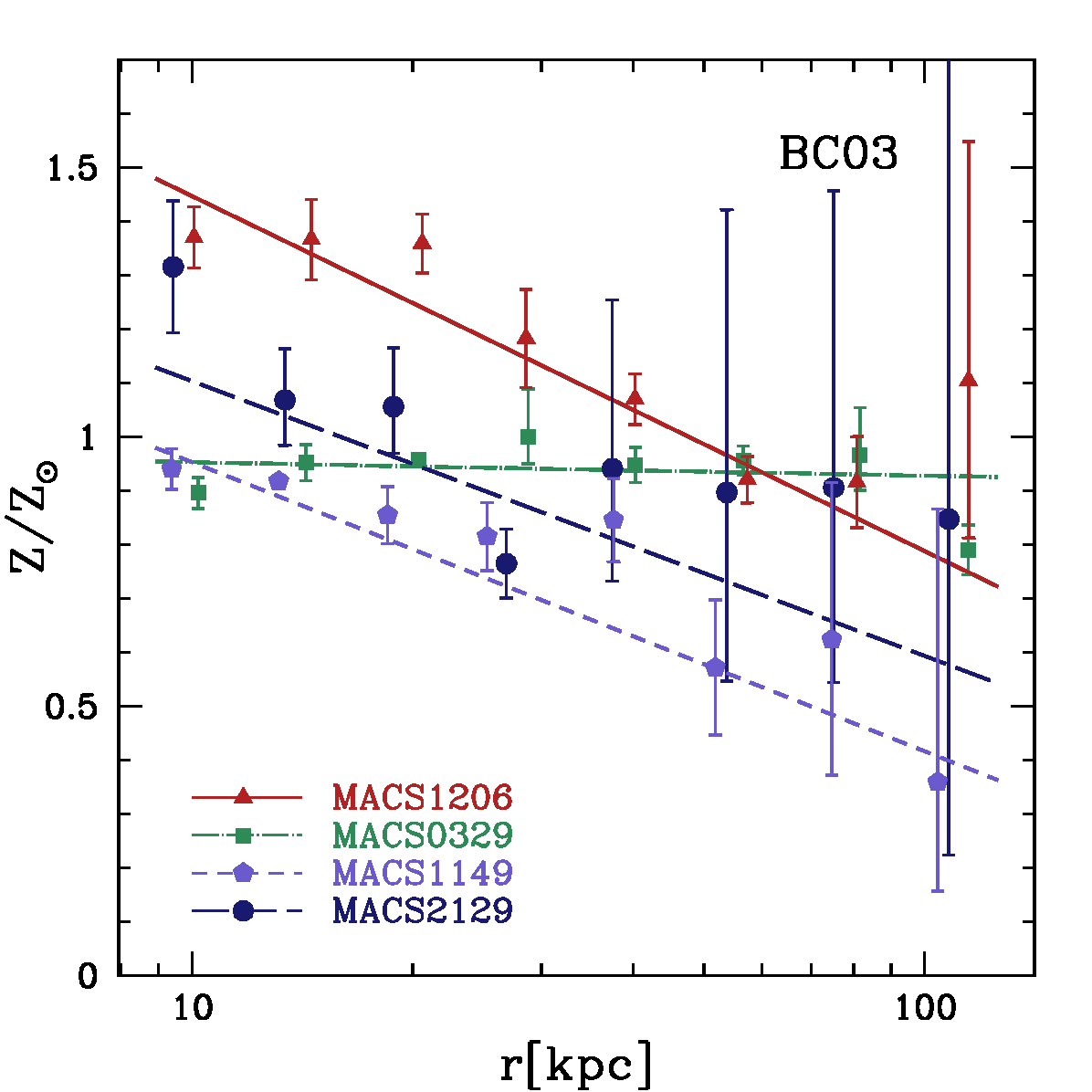}
\caption{The interpolated metallicity profiles of the BCG+ICL produced using the BC03 model set for \zform=3, over-plotted with their best fit linear relation to illustrate the similarity between clusters. The large points are from the metallicity profiles produced with the dlog(r[kpc])=0.15 colour profiles and the best fit relation shown is the result of a weighted linear fit to the dlog(r[kpc])=0.01 profiles. All four clusters show \til \Zsolar\ metallicities within 10 kpc and either remain at \Zsolar\ or dip to sub-solar metallicities by 110 kpc. \MACSohthree\ is the only cluster to show a flat metallicity profile, which is likely due in part to the cluster's unrelaxed dynamical state and contribution of ICL from the massive elliptical galaxy located at a projected 40" north-west from the central BCG.}
\label{fig:overmets}
\end{figure}

We fit the metallicity profile of each cluster with a weighted linear fit between 10$<$r$<$110 kpc and find that the metallicity profile of \MACSohthree\ is consistent with being flat, while the other three clusters have a range of metallicity gradients of -0.510$<$\metgrad$<$-0.659.
In Figure \ref{fig:overmets}, we overlay the weighted linear fits on the interpolated metallicity points for a BC03 model with \zform=3. 
Using the results from the BC03 model as a fiducial, we see that at radii $<$10 kpc all clusters have abundances $\ge$\Zsolar\ and all but \MACSohthree\ dip to solar abundance and below at larger cluster radii. 

At low redshifts \cite{Loubser2009} find spectroscopically measured BCG metallicities of \til2 \Zsolar. Looking to our results, and using BC03 as a fiducial, we find that all four clusters have similar BCG metallicity of \til1-1.4 \Zsolar\ for the BC03 model. 
Referring to Figure \ref{fig:met_panel} we see that the M05 model set predicts metallicities closest to the observed range of \cite{Loubser2009} and \cite{Lidman2012a}.

\begin{figure}
\centering
\includegraphics[width=0.5\textwidth]{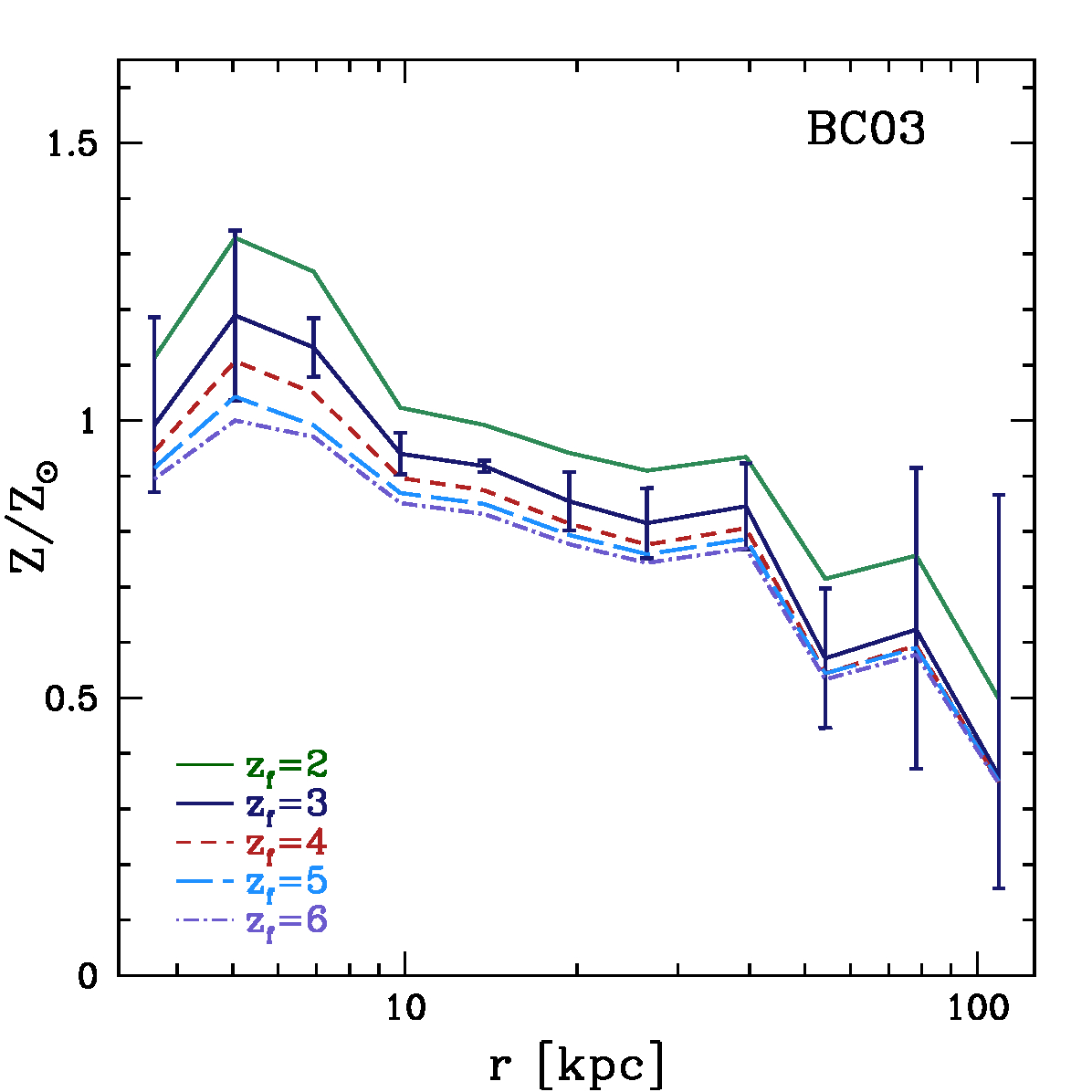}
\caption{Effect of formation redshift on the metallicity interpolated from the F110W-F160W colour of the BCG+ICL of \MACSeleven\ using the BC03 model set.  We produce interpolated metallicity profiles using \zform=2-6, omitting errors bar on all but the \zform=3 line for clarity. Changing the formation redshift from \zform=2 to \zform=6 affects the metallicity by a maximum of only \til0.2 dex and, most importantly, preserves downward trend in metallicity with radius. This suggests that, despite the uncertainty in the age of the ICL, the general trends in metallicity that we observe are robust.}
\label{fig:zfdiff}
\end{figure}

While there are many free parameters for each model set, the most significant for our analysis is the formation redshift. 
Our baseline assumption of \zform=3 is in agreement with the model predictions and observations \citep{sommer, Puchwein2010, Williams2007a, Loubser2009, Coccato2011a, Melnick2012}. 
We present the metallicity profile of \MACSeleven\ as a function of formation redshift for a BC03 model (Figure \ref{fig:zfdiff}) to illustrate how the choice of formation redshift affects the final metallicity gradient.
From \zform=2 to \zform=6 we see a \til0.2 Z/\Zsolar\ decrease in the inferred metallicity.
However, the downward trend in metallicity with radius remains, regardless of which formation redshift is used.
This suggests that, despite the uncertainty in the ICL age, the trends exhibited by our metallicity profiles are robust and that only when considering absolute metallicity does model choice play a noticeable role in the ICL metallicity.

%

\section{Notes on Individual Clusters}
\label{sec:indiv}

Two of our clusters, \MACSohthree\ and \MACStwelve, are included in the list of eight CLASH clusters with possible substructure \citep{Postman2012a}.
The colour and metallicity gradients of the ICL of \MACSohthree\ may be flattened due to light that is contributed from a second giant elliptical galaxy located 40$^{\prime\prime}$ north-west from the central BCG. 
To test this supposition we mask the entire half of the cluster containing the second giant elliptical and compare the colour profiles before and after the additional masking. We find the colour profiles to be consistent, suggesting that if the two galaxies have interacted and deposited stellar populations into the ICL, those additions have been well-mixed with the rest of the ICL. 
\cite{Postman2012a} note that \MACSohthree\ was initially classified as dynamically relaxed by \cite{A08} and \cite{Schmidt2007a} but that later \cite{Maughan08} found evidence of substructure. 
The possible dynamical history that this substructure suggests could have flattened the stellar population gradient of the BCG+ICL component and produced the observed nearly constant colour and metallicity of \MACSohthree's ICL.	

The dynamical state of \MACStwelve\ remains unclear.
Many signs point to the cluster being dynamically relaxed: the optical position of the BCG and X-Ray peak are coincident \citep{Postman2012a}, it has a symmetric X-Ray morphology \citep{gilmour}, the dark matter, BCG, and X-Ray centroids are aligned \citep{Umetsu2012a, Biviano2013a} and show no signs of recent merging activity. 
However, there are multiple signatures indicative of recent merger activity. 
These signposts include a high velocity dispersion ($>$1500 \kms) \citep{gilmour, Biviano2013a}, high X-Ray luminosity and temperature \citep{Ebeling2010a, Postman2012a}. 
Additionally, \MACStwelve\ displays a low surface brightness structure (down to \til27 \sbu) extending from the BCG roughly 400 kpc to the north-west (See Figure \ref{fig:m1206image}, \cite{Presotto2014a,Umetsu2012a}).
Such a distinct feature, which is aligned with the BCG position angle and is traced by in-falling galaxies \citep{Presotto2014a}, suggests recent dynamic activity.
	
\section{Discussion}
\label{sec:discuss}
\subsection{ICL Progenitor Populations}

Here we consider the possible mechanisms of formation of the ICL, describing how the colour and metallicity gradients can be used as diagnostics. 

\subsubsection{Formation Mechanisms}
\label{sec:mech}
The exact mechanisms of ICL formation remain poorly constrained. Four distinct mechanisms have been proposed to explain the formation of the ICL:

\begin{enumerate}[(1)]
    \item \emph{Shredding of dwarf galaxies.}  As low-density galaxies fall into the cluster, they can be tidally disrupted by the cluster potential or by encounters with cluster member galaxies. Their stars become unbound, responding to the cluster potential, and form the intracluster starlight. The most metal-poor dwarfs, which are also the lowest mass due to the mass-metallicity relation, are disrupted at larger cluster radii than more massive satellites, thus forming a gradient in colour and metallicity with cluster radius \citep{Rudick2009, Melnick2012,conroy07}.  

	\item \emph{Tidal stripping.} The tidal stripping of \Lstar\ galaxies from interactions with other cluster members or the cluster potential can contribute to the build-up of the ICL. Because \Lstar\ galaxies have radial metallicity gradients \citep{Zaritsky1994a,La-Barber2012a}, the tidal forces of the cluster potential will liberate progressively more tightly bound (redder) stars as a galaxy approaches the cluster centre. The stars added to the ICL will therefore be somewhat more metal-rich at smaller cluster radii. See \cite{Melnick2012, Rudick2009, Rudick2006, Rudick2010, Puchwein2010,murante07,conroy07}.

	\item \emph{Violent relaxation during mergers.} The ICL may be built up via liberation of stars from galaxies undergoing violent mergers with the centrally dominant cluster galaxy. It has been suggested that 30-50\% of stars of the merging body become unbound and end up in the ICL \citep{murante07,conroy07,Lidman2012a}. Further, major mergers serve to flatten any existing stellar population gradient \citep{White1980a, Kobayashi2004a, Di-Matteo2009a,La-Barber2012a, Eigenthaler2013a}. Thus, if major mergers are a dominant process of ICL formation, then the BCG+ICL colour and metallicity gradients are expected to be shallow or near-constant. The metallicity of the ICL will depend on the masses of the galaxies that interact with the BCG. The more often massive galaxies merge with the BCG, the more metal-rich the intracluster stellar population will be. 

	\item \emph{In situ \emph{formation.}} Finally, a portion of the intracluster stellar population may arise from in situ star formation \citep{Puchwein2010}, wherein clumps of gas stripped from cluster galaxies remain cool and compact enough to form new stars. While the impact of ongoing star formation on the colour and metallicity gradient of the ICL+BCG is unclear, the process would change the mean age of the ICL stellar population, which must be taken into account to successfully interpret any observed colour trends. Little observational evidence exists suggesting that in situ stars are a significant population in the ICL. For example, using synthetic models for the ICL, \cite{Melnick2012} find that only 1\% of the ICL consists of in situ stars in their study of a z=0.29 cluster. The lack of Type Ia intracluster stellar supernovae in z\til0.1 clusters \citep{Sand2011a} further suggests that in situ star formation is not the dominant mechanism of ICL formation.
\end{enumerate}

Current simulations present a range of predictions for the relative importance of these mechanisms. 
Some suggest that disruption of satellites and tidal stripping (mechanisms (1) \& (2) above) may contribute significantly to the ICL stellar population \citep{purcell07, Stanghelli2006a, Rudick2009, Contini2013a, Laporte2013a}, whereas others predict that the majority of the ICL is built up from violent mergers of cluster members with the centrally dominant galaxy (mechanism (3), \cite{murante07, conroy07}). 
The results of \cite{Contini2013a} favour the scenario in which stellar stripping and disruption of satellite galaxies are the primary mechanisms by which the ICL forms. 
\cite{Contini2013a} use updated semi-analytical models based upon \cite{DLB2007} to specifically address how the ICL forms and which progenitor population contributes most to the ICL stellar population. 
They suggest that mergers contribute relatively little to the ICL, in contrast with the results of \cite{murante07}, who argue that tidally-stripped stars represent less than 10\% of the total ICL. 

Using their tidal stripping and satellite disruption models, \cite{Contini2013a} find that 26\% of the ICL starlight results from galaxies with stellar masses in the range of 10$^{10.75}-$10$^{11.25}$ \Msun\ and that 68\% of ICL starlight comes from galaxies with M$_*>$10$^{10.5}$ \Msun. 
Intracluster stars from low-mass galaxies (M$_*<10^9$ \Msun) contribute negligibly to the total ICL population. 
This, they argue, is a consequence of low-mass satellites spending more time at the cluster outskirts, where tidal stripping and disruption are less efficient. 
This picture fits with observed cases of mass segregation in clusters e.g. \cite{Andreon2002a, Biviano2002a, Presotto2012a}.

Using the observed colour and metallicity gradients of the ICL presented here, we comment on the likely importance of the various mechanisms of ICL formation and on the dominant intracluster stellar progenitor population.


\subsection{ICL Assembly History}

Over the redshift range of our sample (z=0.44-0.57) the evolutionary models of \cite{Contini2013a} predict that the ICL grows by \til40\%, increasing from \til35 to 50\% of its present day mass for their "disruption" and "tidal" models (See their Figure 6). 
If we assume that 1) our clusters are different versions of the same evolving cluster with an age spread corresponding
to the redshift interval of our sample and 2) the ICL mass growth corresponds directly to a rise in surface brightness, then both of their models predict a scatter of 0.15 \sbu\ among the surface brightness profiles.  As we have neglected all sources of scatter other than cluster redshift, the model estimates are strict lower limits.

The observed profiles of the four clusters in our sample exhibit a mean scatter of 0.27 \sbu\ between 10<r<110 kpc. 
At these radii, the physical scatter dominates over observational uncertainties.  
Furthermore, as noted in Section \ref{sec:sbu}, a simple mass rescaling does not reduce the measured scatter.
Therefore, for the range of cluster mass probed here, we find no dependence of the normalization of the ICL profile on cluster mass.

We next estimate the contribution of observational errors in the surface brightness profiles to the scatter. 
Quantifying the error in the scatter empirically from the systematic variations among our profiles is difficult due to the small sample size of our current sample.  
With this caveat in mind, we nevertheless use our data to estimate an uncertainty in the RMS values of \til0.05 \sbu, plotting an error bar for reference in the rightmost point in Figure 8. 
Many points are more than 1$\sigma$ above the model RMS of 0.15 \sbu, and all lie above it. 
We conclude that the measured scatter does indeed exceed what is predicted from simple mass growth, suggesting a wider range of assembly processes, epochs, and/or ICL content.  
We leave a more detailed discussion for future papers that will contain our larger, complete sample of clusters.

None of the existing theoretical studies publish the scatter among surface brightness (or colour) profiles as a function of redshift or radius.
We advocate for direct comparisons between models and theoretical predictions for the scatter in ICL build-up among clusters, particularly as the observational work expands to larger samples spanning a wider redshift range. 
Our future work with the full set of CLASH clusters \citep{Postman2012a} and a sample of galaxy groups (HST Program 12575) is intended to provide this type of benchmark, as it will use systems across a range of redshifts with masses that reflect the expected hierarchical build-up of clusters.


\subsubsection{Colour and Metallicity Gradients as Diagnostics}
\label{sec:cnm}

\begin{table*}
\centering
\caption{ICL Colour, Metallicity Gradients and Luminosity}
\begin{tabular}{c c c c c c c}
\hline\hline
Cluster & z & \colorgrad & \metgrad & L(r$\le$10 kpc) & L(r$\le$110 kpc) & L(10 $\le$ r(kpc) $\le$ 110)  \\ 
& & [\sbu\ log(kpc)$^{-1}$] & [log(kpc)$^{-1}$] & [L$_*$] & [L$_*$] & [L$_*$] \\[0.5ex]
\hline
\MACStwelve 	& 0.44 & -0.091 $\pm$ 0.006  & -0.659 $\pm$ 0.035 & 1.11 & 5.76  & 4.65 \\
\MACSohthree 	& 0.45 &  0.000 $\pm$ 0.004  & -0.025 $\pm$ 0.018 & 1.43 & 8.04  & 6.61 \\
\MACSeleven 	& 0.54 & -0.101 $\pm$ 0.014  & -0.537 $\pm$ 0.062 & 1.75 & 10.26 & 8.51 \\
\MACStwentyone 	& 0.57 & -0.063 $\pm$ 0.014  & -0.510 $\pm$ 0.105 & 1.47 & 6.85  & 5.38 \\
\hline

\end{tabular}
\label{table:fact}
\end{table*}

The colour and metallicity of the ICL reflects the dominant progenitor population, encoding the formation history of the cluster in the ICL \citep{Zibetti2005, Willman2004a, Murante2004a, sommer}. 
\cite{purcell07} predicts that clusters with a halo mass greater than $10^{13.5} M_\odot$ will have an ICL of solar metallicity and the simulations of \cite{sommer} show ICL colours consistent with a progenitor population of sub-\Lstar, E, and S0 galaxies. 

The colour gradients of the ICL in our clusters significantly trend toward bluer colour at larger cluster radii, except for \MACSohthree, which has a flatter profile (See Figure \ref{fig:allover}). 
\cite{Zibetti2005} similarly find a blue-ward gradient in their analysis of stacked SDSS clusters and \cite{Montes2014} show that the Frontier Fields cluster Abell 2744 has a bluening colour profile in both g-r and i-J.
These observations strongly favor the formation mechanisms that produce distinct metallicity gradients: tidal stripping and disruption of the \Lstar\ and dwarf galaxy populations.

In addition to the \blue$-$\red\ colour gradients, we also use the absolute metallicities and metallicity gradients of the ICL in these clusters to constrain the progenitor population and formation mechanism of the ICL (See Figure \ref{fig:overmets}). 
To this end we perform weighted linear fits to the metallicity profiles from the BC03 model for 10$<$r$<$110 kpc. 
\MACStwelve, \MACSeleven, and \MACStwentyone\ all exhibit steep metallicity gradients of, on average, \metgrad\til\ -0.6 log(kpc)$^{-1}$. 
Including the shallower profile of \MACSohthree\ results in an ensemble average gradient of \metgrad\til\ -0.4 (see Table \ref{table:fact} \& Figure \ref{fig:met_panel}). 

While major mergers may be the dominant mechanism by which the BCG grows \citep{Lidman2013a}, we conclude that they are less important in the build-up of ICL outside of the BCG because of the significant observed colour and metallicity gradients. 
Between 10 and 110 kpc we find total ICL luminosities of 4-8 $L_*$ (Table \ref{table:fact}). 
If the ICL is built-up solely from stars ejected during accretion of \Lstar\ galaxies by the BCG, and \til20\% of the stars of the merging galaxy are deposited in the ICL (as in \cite{murante07, Contini2013a}) then 20-40 merging events must occur.
This number of merging events is 10 times greater than that found in models of BCG evolution since z  =1 \citep{Lidman2013a}. 
Further, this number of major mergers is enough to significantly erode any underlying stellar population gradient \citep{Kobayashi2004a, Di-Matteo2009a}.
Additionally, the higher central concentration of ICL relative to the galaxy distribution is consistent with a formation mechanism that is more efficient deeper in the cluster potential well e.g. tidal interactions \citep{Zibetti2005}. 

Further, the near-solar metallicities imply that the ICL cannot be formed solely from low-metallicity dwarf galaxies. 
This result echoes the conclusions of \cite{Contini2013a}, whose models indicate that galaxies with stellar masses of \til 10$^{11} M_\odot$ are the primary contributors to the ICL. 
They frame their conclusion in terms of dynamical friction: massive satellite galaxies experience the greatest dynamical friction, causing them to quickly fall into the central region of the cluster where they can be tidally stripped and disrupted. 
If the observed 4-8 \Lstar\ of ICL luminosity formed solely from dwarf disruption, then hundreds of dwarfs would be required.
The faint end of the luminosity function must have evolved from a slope at least as steep as $\alpha=-1.8$ to the flatter values observed today (-0.84 \cite{Lin2004a}) in order for the difference in integrated luminosity between the past and present luminosity functions to be greater than 4\Lstar, the minimum observed ICL luminosity.
However, the faint-end slope has been shown to have little or no evolution since z\til1.3 \citep{Mancone2012a} and possibly since z=3.5 \citep{Stefanon2013a}.

\cite{Zibetti2005} and \cite{KrickI} find that the optical colours of the ICL are well-matched by the cluster galaxy colour distribution, suggesting that the ICL results in part from stars stripped and disrupted from red sequence galaxies. 
The observed \blue$-$\red\ colour range of our clusters also closely matches their red sequence colours, in general spanning colours equal to the cluster's red sequence down to \Mstar+2 mag. 
This further supports the idea that the progenitor population of the ICL is largely from \Lstar\ galaxies. 

While it is possible, though unlikely, for a colour and metallicity gradient to develop via in situ formation, the metallicity of the resulting intracluster stars would likely be similar to the ICM metallicity of \til 0.3 \Zsolar\ \citep{Edge1991a, Sivananda2009a}. 
In two of our clusters  (\MACSeleven\ and \MACStwentyone), the metallicity gradients reach down to \til0.3 \Zsolar\ by 110 kpc. 
However, those metallicity profiles are produced under the assumption of an old stellar population with \zform=3. 
Assuming instead that the ICL is developed via in situ star formation and that the predicted ICL formation epoch of z$<$1 is correct \citep{murante07,DLB2007}, the formation redshift of the intracluster stars would be around \zform=1. 
For example, a stellar population at the redshift of \MACSeleven\ (z=0.544) that formed at z=1 with a metallicity of Z=0.4 \Zsolar\ would have a \blue$-$\red\ colour of 0.21, as found using a BC03 SSP model with a burst of star formation with $\tau$=1 Gyr e-folding time and a Chabrier IMF. 
This is inconsistent with the observed \blue$-$\red\ colour profiles, where, for example, \MACSeleven\ has a \blue$-$\red\ colour of 0.33 at r=110 kpc. 
If the ICL was formed in large part by in situ star formation, we would expect to see much bluer colour profiles. 


\section{Conclusions}
\label{sec:concl}
In this paper we have presented initial results from a comprehensive \emph{HST} program to quantify the properties of intracluster light in a statistical sample of clusters and groups at intermediate redshift.
Using the spectacular spatial resolution and depth of HST, we have observed surface brightness and colour profiles for a unique sample of four galaxy clusters at redshifts 0.44$<$z$<$0.57, ranging in mass from 0.6 to 2.6 $\times$10$^{15} M_\odot$ .
\begin{enumerate}[(1)]
\item The F160W surface brightness profiles of our sample have an average RMS scatter between 10$<$r$<$110 kpc of 0.24 \sbu (See Figure \ref{fig:rmsboth}).
This scatter is physical -- it exceeds the observational errors, straightforward expectations from the range of cluster masses in our sample, and predictions based on published evolutionary models for the variance attributable to the redshift span of our sample.
We associate the additional scatter with differences in ICL assembly process, formation epoch, and/or ICL content.

\item Three out of four of our clusters have intracluster stellar populations that get progressively bluer with increasing radius, with a mean colour gradient of -0.085$\pm$0.021 mag log(kpc)$^{-1}$ (-0.064$\pm$0.021 mag log(kpc)$^{-1}$ including the flat profile of \MACSohthree). 
If we interpret this colour gradient as a metallicity gradient for an old stellar population, the metallicity decreases from high values ($\ga$\Zsolar) in the central region where the brightest cluster galaxy (BCG) dominates ($\la$10 kpc) to subsolar at the largest radii. 
The average metallicity gradient for the 3 clusters with negative colour gradients is \metgrad=-0.57$\pm$0.13 log(kpc)$^{-1}$ (-0.43$\pm$0.13 log(kpc)$^{-1}$ for all 4 clusters). 
Such negative colour gradients can arise from tidal stripping of L* galaxies and/or the disruption of dwarf galaxies, but not major mergers with the BCG.

 \item The ICL at \til110 kpc has a colour comparable to \Mstar$+2$ red sequence galaxies, suggesting that out to 110 kpc the ICL is dominated by stars liberated from galaxies with $L>0.16$\Lstar. 
The colours (and equivalently high inferred metallicities) disfavour disruption of even lower luminosity dwarf galaxies as the dominant formation mechanism for the ICL. 

\item For 10$<$r$<$110 kpc, the luminosity of the ICL for these clusters is 4-8 \Lstar. 
For dwarf disruption to be the sole source of ICL, more substantial evolution of the faint end of the luminosity function than observed  would be required.
Thus, the total luminosity of the ICL also argues against dwarf disruption as the dominant progenitor population for the ICL.  
Further, the observed ICL luminosity also disfavours major mergers with the BCG as the dominant formation mechanism of the ICL as a merger model origin for the ICL requires 10 times more mergers than predicted in BCG formation models \citep{Lidman2012a}.
\end{enumerate}

To summarize, using this sample of CLASH clusters we set empirical limits on the amount of scatter in ICL surface brightness profiles at z=0.5 for theoretical models and constrain the progenitor population and formation mechanism of the ICL. 
The results from the absolute colour, colour gradients, and total luminosity of the ICL, as enumerated above, are suggestive of a formation history for these clusters in which the ICL is built-up by the stripping of >0.2\Lstar\ galaxies, and disfavour significant contribution to the ICL by dwarf disruption or major mergers with the BCG.
Our future work will include analysis of the remaining clusters from the CLASH sample, as well as eight groups from our HST-GO program (Program \#12575), expanding the mass range of the cluster environments. 
This larger mass range, combined with the more-extended redshift baseline, will allow us to trace the build-up of the ICL over the past 5 Gyr. In addition, we will use this high- fidelity, multi-band photometry to structurally disentangle the BCG and ICL.

 \section*{Acknowledgements} 
We acknowledge support from the National Science Foundation through grant NSF-1108957. Support for Program number 12634 was provided by NASA through a grant from the Space Telescope Science Institute, which is operated by the Association of Universities for Research in Astronomy, Incorporated, under NASA contract NAS5-26555.

\bibliographystyle{mn2e}{.bbl}
\bibliography{ICLNov2014}

\end{document}